\documentclass[prb,twocolumn,showpacs,longbibliography,superscriptaddress, groupedaddress]{revtex4-1}

\usepackage{graphicx}

\usepackage{amsmath}
\usepackage{amssymb}
\usepackage{dcolumn}
\usepackage{bm}
\usepackage{wasysym}
\usepackage{dsfont}
\usepackage{enumitem}
\usepackage{hyperref}
\usepackage{braket}

\newcommand{\ii}{{\rm i}}

\newcommand{\cA}{ {\cal A} }

\newcommand{\cC}{ {\cal C} }
\newcommand{\cD}{ {\cal D} }

\newcommand{\cL}{ {\cal L} }

\newcommand{\bt}{{\bf t}}
\newcommand{\sT}{\mathsf{T}}
\newcommand{\tbb}{\tilde{\bf b}}
\newcommand{\bb}{{\bf b}}
\newcommand{\bl}{{\bf l}}
\newcommand{\bx}{{\bf x}}
\newcommand{\mfb}{\mathfrak{b}}

\newcommand{\Udual}{\mathrm{U}(1)_{\rm dual}}
\newcommand{\Uc}{\mathrm{U}(1)_{\rm c}}
\newcommand{\Ur}{\mathrm{U}(1)_{\rm r}}

\newcommand{\SO}{\mathrm{SO}}

\newcommand{\SU}{\mathrm{SU}}
\newcommand{\U}{\mathrm{U}}

\newcommand{\Tr}{\mathrm{Tr}}

\usepackage{xcolor}

\begin{document}

\title{Theory of Topologically Ordered Superfluids in 2+1 Dimensions}
\author{Chao-Ming Jian}
\affiliation{Department of Physics, Cornell University, Ithaca, New York 14853, USA}

\begin{abstract}
Topologically ordered superfluids (TOSFs) support deconfined, gapped anyonic excitations while remaining gapless because of Goldstone modes. We develop a general framework for characterizing their topological data in $(2+1)$-dimensional bosonic systems with spontaneously broken charge $\Uc$ symmetry. Gauging $\Uc$ maps a TOSF to a gapped topological order enriched by a dual $\Udual$ symmetry with vanishing Hall conductance, while gauging $\Udual$ recovers the TOSF. This two-way mapping allows the topological data of the TOSF, including the condensate charge, deconfined anyon content, and topological properties of vortices, to be systematically encoded using the algebraic framework of symmetry-enriched topological orders. In particular, superfluid vortices share the topological properties of flux defects associated with the residual discrete symmetry in the gapped anyon sector. We develop complementary algebraic and field-theoretic formulations of this framework, with the latter also suggesting a natural transition to a gapped topologically ordered phase. Additionally, using parton constructions, we present examples of TOSFs with non-Abelian vortices in bosonic systems.
\end{abstract}

\date{\today}

\maketitle

\section{Introduction}
For $2+1$D gapped quantum systems at zero temperature, the fundamental data of the topological order (TO), including the anyon content, their fusion rules and braiding statistics, can be systematically characterized within the algebraic framework of unitary modular tensor categories\cite{Kitaev2006}. In the presence of a global symmetry, the TO is further enriched by fractionalized symmetry transformations of the anyons and the topological properties of symmetry defects. These additional data can be organized using $G$-crossed extensions of the unitary modular tensor category framework\cite{EtingofNOM2009,SET,TeoSET2015,TarantinoSET2016}. While these frameworks are well established for gapped systems in 2+1D, they do not immediately generalize to generic gapless systems.

An interesting class of gapless systems consists of superfluids in which gapless Goldstone modes coexist with deconfined, gapped anyonic excitations. Such phases and their superconducting counterparts have been explored in earlier work\cite{WenZee1992NeutralSF,SenthilFisher2000ElecFrac,Demler2002,Paramekanti2004,TangWen2013,BondersonNayak2013QuasiTopo,Mong2014,Gazit2017}. More recently, a rich set of examples has been found by deforming or doping fractionalized quantum states\cite{HanWang2026AnyonSF,WangZhang2026AnyonSF,Song20254e,ShiSenthil2025NonAbelianSC,ShiZhangSenthil2025,GaoWu2025,Gao2026,ShiHan4e2026,ShiSenthil2025}. Notably, Ref.~\onlinecite{ShiHan4e2026} identifies a charge-$4e$ topological superconductor that supports not only gapped anyonic excitations but also vortices hosting non-Abelian parafermion zero modes, which can be used for topological quantum computation. 

In the following, we refer to superfluids and superconductors that support deconfined, gapped anyonic excitations as topologically ordered superfluids (TOSFs). Due to the gapless Goldstone modes associated with the spontaneous breaking of a global $\U(1)$ symmetry, the TOSFs are not directly described by the established category-theoretic approaches for TOs. A new systematic framework is needed for characterizing their topological data, including both the deconfined anyons and the topological properties of vortices. Such a framework would extend the systematic study of topological phases of matter beyond fully gapped systems and help identify potential schemes for quantum computation using TOSFs.

In this work, we develop such a framework by combining the algebraic characterization of symmetry-enriched TOs with the framework for gauging $\U(1)$ symmetry in $2+1$D TOs\cite{GaugingU1}. We focus on $2+1$D bosonic systems with a spontaneously broken global charge $\Uc$ symmetry. We show that gauging $\Uc$ maps a gapless TOSF $\cC$ to a gapped TO $\cD$ enriched by a dual (unbroken) $\Udual$ symmetry, whose topological data can be described within the established algebraic framework for symmetry-enriched TOs. Conversely, gauging $\Udual$ in $\cD$ recovers the TOSF $\cC$. Through this two-way mapping via gauging, the charge of the condensate, the deconfined gapped anyons, and the topological properties of the vortices in the TOSF $\cC$ are systematically encoded by the corresponding $\Udual$-enriched TO $\cD$.

We first provide an algebraic formulation of this framework and then develop a complementary field-theoretic description. A key result identifies the topological degrees of freedom carried by superfluid vortices with those of discrete symmetry defects acting on the gapped anyon sector of the TOSF. This correspondence was also obtained in Ref.~\onlinecite{Else2021TopoGoldstone} through an argument based on continuously parametrized families of gapped TOs. Our approach places the correspondence within a unified framework that also encodes the condensate charge and the deconfined anyon content. The complementary field-theoretic formulation corroborates the algebraic results and provides connections to microscopic constructions and the study of possible phase transitions. In addition to introducing this general framework using algebraic and field-theoretic approaches, we also provide novel examples of TOSFs with non-Abelian vortices using parton constructions. 

The rest of the paper is organized as follows. In Sec.~\ref{sec:GeneralFormalism}, we establish the two-way mapping between a general TOSF $\cC$ and a $\Udual$-enriched gapped TO $\cD$ via gauging. We use the algebraic description of $\cD$ to develop a general framework for characterizing the topological data of $\cC$. In Sec.~\ref{sec:FieldTheory}, we present a complementary field-theoretic formulation, treating the TOSFs encoded by Abelian and non-Abelian TOs separately. We also provide concrete examples of TOSFs with non-Abelian vortices. In Sec.~\ref{sec:conclusion}, we summarize our results and discuss future directions.

\section{General Algebraic Framework}
\label{sec:GeneralFormalism}

In this section, we introduce a general algebraic framework for characterizing the topological data of a TOSF $\cC$ in a 2+1D bosonic system. This TOSF $\cC$ spontaenously breaks a global 0-form charge symmetry $\Uc$. We formally denote the Lagrangian of this phase by $\cL_\cC[A]$, which includes the coupling of $\cal C$ to the 1-form background gauge field $A$ associated with the $\Uc$ symmetry. Our goal is to systematically characterize the anyon contents, condensed objects, and vortices in the TOSF $\cC$.

Consider gauging the global $\Uc$ symmetry of $\cal C$ by promoting the background $\Uc$ gauge field $A$ to a dynamical one, denoted by $\alpha$. The fluctuating gauge field $\alpha$ absorbs the Goldstone mode of $\cal C$ and, consequently, maps the TOSF phase $\cal C$ to a fully gapped TO $\cal D$ in 2+1D. Throughout the paper, the term TO refers to the low-energy anyon model or the topological quantum field theory description of a gapped system. The TO $\cal D$ is naturally enriched by a dual global 0-form $\U(1)_{\rm dual}$ symmetry that couples to monopoles of the dynamical gauge field $\alpha$. This gauging process yields the following Lagrangian for the TO $\cal D$:
\begin{align}
    \cL_{\cal D}[B]=\cL_{\cal C}[\alpha]+\frac{1}{2\pi}B d\alpha,
    \label{eq:LagrangianD}
\end{align}
where $B$ is the 1-form background gauge field associated with the $\Udual$ symmetry.

An important observation is that, upon further gauging the $\U(1)_{\rm dual}$ symmetry of the TO $\cal D$ by promoting $B$ to a dynamical gauge field $\beta$, we recover the TOSF phase $\cal C$. The $\U(1)$ symmetry that couples to $\frac{d\beta}{2\pi}$ is identified with the original charge $\Uc$ symmetry of $\cal C$ (up to a redefinition of the sign of the charges). This observation is evident from the Lagrangian formulation:
\begin{align}
   \cL_{\cal D}[\beta]-\frac{1}{2\pi}A d\beta
   \longrightarrow \cL_{\cal C}[A],
\end{align}
where the left-hand side is the Lagrangian after $\U(1)_{\rm dual}$ is gauged, while the right-hand side is obtained by integrating out the dynamical gauge field $\beta$, which sets $\alpha=A$. This observation indicates that we can obtain an algebraic characterization of the 2+1D TOSF $\cal C$ by combining the category-theoretic description of the symmetry-enriched TO $\cal D$\cite{SET,TarantinoSET2016,TeoSET2015} with the algebraic theory of $\U(1)$ gauging in 2+1D topological orders.\cite{GaugingU1,GaugingLieGroup}

We now characterize the $\U(1)_{\rm dual}$-enriched TO $\cal D$. The enrichment of a generic 2+1D TO by an unbroken $\U(1)$ symmetry is fully determined by two pieces of data\cite{SET,TarantinoSET2016,TeoSET2015,GaugingU1}: (1) the Abelian anyon generated by the $2\pi$ symmetry flux, called the vison, and (2) the Hall conductance $\sigma_{xy}$\footnote{The sign convention for the Hall conductance $\sigma_{xy}$ is fixed by the response Lagrangian $\cL_{\rm response} = -\frac{\sigma_{xy}}{4\pi}BdB$, where $B$ is a background $\U(1)$ gauge field.}. In particular, the fractional $\U(1)$ charges of all the anyons are determined by their braiding statistics with the vison, while the Hall conductance $\sigma_{xy}$ is constrained by the vison's topological spin, $\theta=e^{\ii\pi\sigma_{xy}}$. For the $\Udual$-enriched TO $\cal D$, we denote the corresponding vison by $v$. Its order is denoted by $n$; namely, $v^n$ is identified with the trivial anyon in $\cal D$. Based on Eq. \eqref{eq:LagrangianD}, $v$ carries unit charge under the dynamical gauge field $\alpha$. The topological phase $\cal D$ must have $\sigma_{xy}=0$ for the $\U(1)_{\rm dual}$ response. This is because, if $\sigma_{xy}$ were nonzero, gauging $\U(1)_{\rm dual}$ would result in a $\U(1)$-symmetric gapped TO with Hall conductance $-\sigma_{xy}^{-1}$,\cite{GaugingU1} rather than the TOSF $\cal C$ with spontaneously broken $\Uc$ symmetry. The vanishing $\sigma_{xy}$ implies that the vison $v\in \cD$ has bosonic self-statistics.

The next step is to gauge the $\U(1)_{\rm dual}$ symmetry of $\cal D$, thereby recovering the TOSF $\cal C$. Let us first ignore the original charge $\Uc$ symmetry, which couples to $d\beta/2\pi$ in this gauging process, or temporarily treat it as explicitly broken. With $\sigma_{xy}=0$, this gauging amounts to condensing the vison $v\in\cD$, resulting in the TO ${\cal D}/{\cal A}$.\cite{GaugingU1} Here, ${\cal A}=\{1,v,\ldots,v^{n-1}\}$ is the condensable algebra generated by the vison $v$. This anyon condensation\cite{BaisSlingerland2009} can equivalently be described as gauging a $\mathbb{Z}_n^{(1)}$ 1-form symmetry\footnote{In our notation, the superscript ``${}^{(k)}$'' emphasizes that the symmetry is a $k$-form symmetry. The $\U(1)$ symmetries in this work are all 0-form symmetries, so we suppress the superscript.} of $\cal D$ generated by the worldline of the vison $v$.\cite{Gaiotto2015,Moore:1989yh} The TO ${\cal D}/{\cal A}$ captures the gapped anyonic excitations in the zero-vorticity sector of the TOSF $\cal C$. 

\begin{figure*}
    \centering
    \includegraphics[width=0.85\linewidth]{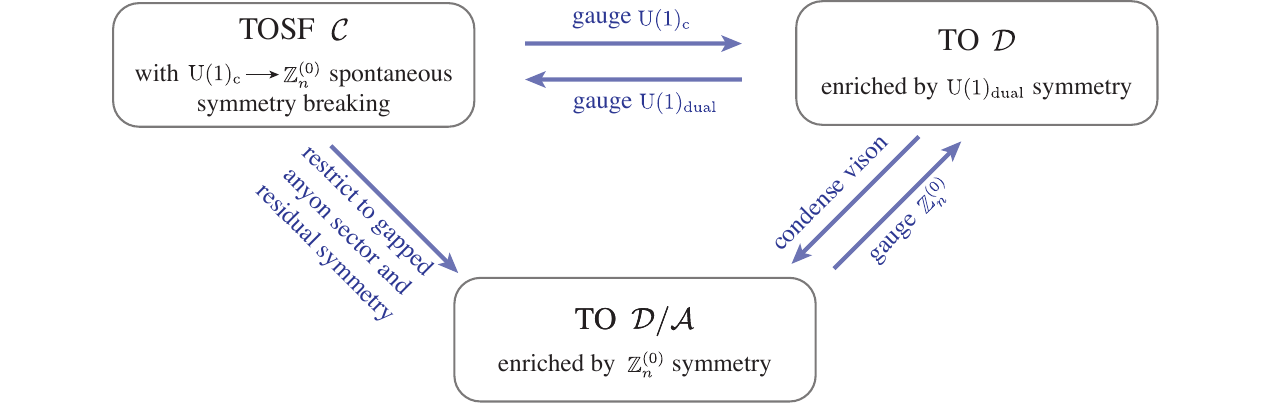}
    \caption{Relations between the TOSF $\cC$, its deconfined anyon sector $\cD/\cA$, and the $\Udual$-enriched TO $\cD$.}
    \label{fig:TOWeb}
\end{figure*}

We now examine the consequences of the original $\Uc$ symmetry in this ``anyon condensate.'' Recall that the vison $v\in\cal D$ carries unit charge under the dynamical gauge field $\alpha$, which is pinned to the background field $A$ upon gauging $\U(1)_{\rm dual}$. The object $v^n$ is the local condensed object in $\cal C$ with the smallest charge under $A$. Therefore, the TOSF $\cal C$ is a charge-$n$ condensate that spontaneously breaks the original $\Uc$ symmetry down to its $\mathbb{Z}^{(0)}_n$ subgroup. The relations between the TOSF $\cC$, its gapped anyon sector $\cD/\cA$, and the $\Udual$-enriched TO $\cD$ are summarized in Fig. \ref{fig:TOWeb}.

The vortices of the TOSF $\cal C$ exhibit quantized $\Uc$ fluxes, or vorticities, in $\frac{2\pi}{n}\mathbb{Z}$. More interestingly, they can carry topological degrees of freedom. Formally, if we explicitly break the original $\Uc$ symmetry down to its $\mathbb{Z}^{(0)}_n$ subgroup, the Goldstone modes of $\cal C$ are gapped out, and the resulting gapped phase exhibits the TO $\cD/\cA$ enriched by the global $\mathbb{Z}_n^{(0)}$ 0-form symmetry. The elementary vortex of $\cal C$, with flux $\frac{2\pi}{n}$, can be identified with the unit $\mathbb{Z}_n^{(0)}$ symmetry flux defect. In the anyon-condensation picture, the $\mathbb{Z}_n^{(0)}$ symmetry fluxes are descendants of anyons in $\cal D$ that are confined by the condensation of the vison $v$ and can therefore carry nontrivial topological degrees of freedom. The topological properties of these $\mathbb{Z}_n^{(0)}$ fluxes, including their braiding and fusion, can be extracted using the theory of anyon condensation.\cite{BaisSlingerland2009} The full topological data of the $\mathbb{Z}^{(0)}_n$-enriched TO $\cal D/\cA$ and its $\mathbb{Z}_n^{(0)}$ fluxes can be organized following the established mathematical structure of a $G$-crossed modular tensor category.\cite{SET,TarantinoSET2016,TeoSET2015} Returning to the TOSF $\cal C$, where the $\Uc$ symmetry is spontaneously broken rather than explicitly broken down to its $\mathbb{Z}^{(0)}_n$ subgroup, the vortices in $\cal C$ should share the same topological properties as the corresponding $\mathbb{Z}_n^{(0)}$ symmetry fluxes. In Sec. \ref{sec:nonAbelianD}, we will construct examples of TOSFs with vortices carrying non-Abelian topological degrees of freedom. 

It is important to emphasize that although vortices and discrete symmetry defects share the same topological properties, they need not share the same energetic properties. In particular, identifying discrete symmetry fluxes with excitations confined by anyon condensation does not imply that vortices are energetically confined in the TOSF. In a purely 2+1D setting, the vortex interactions are mediated by the gapless Goldstone modes of the TOSF.

We have thus shown that the topological data of the TOSF $\cal C$, including its gapped anyonic excitations, the charges of its condensed local objects, and the topological properties of its vortices, can be fully encoded by the $\Udual$-enriched TO $\cD$ through the gauging of $\Udual$. Thus far, our formulation has been fully algebraic. In Sec. \ref{sec:FieldTheory}, we will provide a complementary formulation based on field-theoretic approaches.

Some remarks on the vortices of the TOSF $\cC$ are in order. We have shown that, in bosonic systems, the topological degrees of freedom carried by a vortex in a general TOSF can be identified with those carried by discrete symmetry fluxes in a 2+1D TO enriched by a discrete symmetry. A similar correspondence between vortices and discrete symmetry defects has recently been observed in examples of topological superconductivity in fermionic systems.\cite{ShiHan4e2026,Gao2026,GaoWu2025} Notably, Ref. \onlinecite{ShiHan4e2026} pointed out that vortices carrying non-Abelian degrees of freedom can be used to achieve universal quantum computation. Our framework suggests that topological quantum computation based on TOSF vortices shares the same underlying topological principles as schemes based on discrete symmetry defects in TOs, which have been studied extensively (see, for example, Refs. \onlinecite{Bombin2010_dislocation,Lindner2012,Meng2012,Clarke2013,YouWen2012,BarkeshliQi2012,Barkeshli2012_genon,YouJianWen2013,Yoder2017_surfacecodetwist,kesselring2018,Teo_2016,SET,TarantinoSET2016,TeoSET2015}). Nevertheless, discrete symmetry defects can be challenging to engineer and control in practical quantum devices. TOSF vortices therefore offer an intriguing route to realizing defect-based topological quantum computation. 

Separately, the relation between vortices in TOSFs and discrete symmetry defects associated with the residual unbroken symmetries was also obtained in Ref. \onlinecite{Else2021TopoGoldstone} using a different algebraic argument based on the classification of continuously parameterized families of TOs. The advantage of our approach is that it provides a streamlined and systematic characterization of all the topological data of a TOSF while naturally admitting a complementary field-theoretic formalism (see Sec. \ref{sec:FieldTheory}) that connects more directly to microscopic constructions.

Having shown that the topological data of the TOSF $\cal C$ can be fully encoded by the $\Udual$-enriched TO $\cD$ through the gauging of $\Udual$, we next ask whether multiple distinct $\Udual$-enriched TOs can correspond to the same TOSF $\cal C$. Here, we assume that stacking the $\Uc$-breaking TOSF $\cal C$ with a $\Uc$-symmetry-protected bosonic integer quantum Hall (BIQH) state\cite{BIQH2013} does not change the topological classification of the system. Interestingly, such stacking does change the $\Udual$-enriched TO obtained by gauging $\Uc$. Suppose that the $\Udual$-enriched TO before stacking is $\cD$. Stacking with a BIQH state of Hall conductance $p\neq 0$ with respect to $\Uc$ induces the change
\begin{align}
    \cD \rightarrow \cD'=
    \frac{\cD \boxtimes \U(1)_{-p}
    \boxtimes \U(1)_{n^2p/l^2}}{\tilde \cA}.
    \label{eq:Dprime}
\end{align}
Here, $\tilde \cA$ denotes the condensable algebra generated by the anyon $(v,-1,n/l)$, which is the composite of the vison $v\in\cD$, the elementary anyon with charge $-1$ in the $\U(1)_{-p}$ sector, and the charge-$n/l$ anyon in the $\U(1)_{n^2p/l^2}$ sector. The notation $\U(1)_{-p}$ and $\U(1)_{n^2p/l^2}$ refers to the topological orders described by the $\U(1)$ Chern--Simons theories at levels $-p$ and $n^2p/l^2$, respectively. Here, $l={\rm gcd}(n,|p|)$ is the greatest common divisor of $n$ and $|p|$. The vison of $\cD'$ is given by $(v,0,0)$. Recall that the Hall conductance $p$ of a BIQH state must be an even integer, i.e., $p\in2\mathbb{Z}$. It is straightforward to verify that $\cD'$ encodes the data of the same TOSF $\cC$ (stacked with a BIQH state). The derivation of Eq. \eqref{eq:Dprime} is provided in App. \ref{app:Dprime}.

A more careful analysis shows that $\cD'$ depends only on $p\mod 2n$. Heuristically, this is because the $\Uc$ symmetry is spontaneously broken down to $\mathbb{Z}_{n}^{(0)}$. When we stack a BIQH state with the TOSF $\cal C$, only the ``remaining topology'' of the BIQH state after the reduction of $\Uc$ to $\mathbb{Z}_{n}^{(0)}$ matters for $\cD'$. Note that $\mathbb{Z}_{n}^{(0)}$-symmetry-protected topological phases in 2+1D are classified by the index ${\cal H}^3(\mathbb{Z}_{n}^{(0)},\U(1))=\mathbb{Z}_{n}$. Upon symmetry reduction, the BIQH state reduces to the class labeled by the index $\frac{p}{2}\mod n$. Since $p$ must be even, $\cD'$ depends only on $p\mod 2n$. A more systematic discussion of how $\cD'$ depends on $p$ is provided in App. \ref{app:Dprime}. We caution that our discussion assumes that stacking the TOSF $\cC$ with a BIQH state leaves its topological class unchanged. However, when $\Uc$ is explicitly broken to its $\mathbb{Z}_n^{(0)}$ subgroup, $\cC$ becomes gapped, and the BIQH state can remain nontrivial as a $\mathbb{Z}_n^{(0)}$-symmetry-protected topological state. Stacking can therefore change the topological class of the resulting gapped system. This raises an interesting question: can stacking with a BIQH state also change the topological class of the TOSF $\cC$ without explicitly breaking $\Uc$? We leave this question for future work.

\section{Field theory formulation and examples} 
\label{sec:FieldTheory}

In this section, we discuss a field-theoretic formulation of TOSFs that complements the algebraic framework in Sec. \ref{sec:GeneralFormalism}. We also introduce examples of TOSFs with non-Abelian vortices.

\subsection{TOSF $\cC$ encoded by Abelian TO $\cD$}

We first focus on the field-theoretic formulation of TOSFs encoded by Abelian $\Udual$-enriched TOs. A generic $\Udual$-enriched Abelian TO $\cD$ in a bosonic system admits a $K$-matrix $\U(1)^N$ Chern--Simons description with the following Lagrangian (in real time or Lorentz signature):
\begin{align}
    \cL_{\cD}[B]
    =\sum_{I,J=1}^N\frac{K_{IJ}}{4\pi}b_I db_J
    +\frac{1}{2\pi}t_I Bdb_I,
\end{align}
where $b_{I=1,2,\ldots,N}$ are $N$ dynamical $\U(1)$ gauge fields, which can be organized into a vector ${\bf b}=(b_1,b_2,\ldots,b_N)^\sT$, and $K$ is an $N\times N$ nondegenerate symmetric integer matrix with even diagonal entries. Here, $B$ is the background gauge field associated with the $\Udual$ symmetry and couples to $\cD$ through the charge vector ${\bf t}=(t_1,t_2,\ldots,t_N)^{\mathsf T}$. The vison $v$ generated by the $2\pi$ flux of $B$ is an anyon whose charges under the dynamical gauge fields $\bf b$ are given by the vector $\bf t$. As discussed in Sec. \ref{sec:GeneralFormalism}, for $\cD$ to encode a TOSF $\cal C$, it must have a vanishing Hall conductance with respect to $\Udual$, namely, $\sigma_{xy}={\bt}^{\sT}K^{-1}{\bt}=0$.

The TOSF $\cC$ is obtained by gauging the $\Udual$ symmetry of $\cD$. Hence, the Lagrangian of $\cC$ is
\begin{align}
  \cL_{\cC}[A]
  =\sum_{I,J=1}^N\frac{K_{IJ}}{4\pi}b_I db_J
  +\frac{1}{2\pi}t_I\beta db_I
  -\frac{1}{2\pi}Ad\beta,  
  \label{eq:LagrangianCK}
\end{align}
where we have promoted the background field $B$ to a dynamical $\U(1)$ gauge field $\beta$ and introduced the background gauge field $A$ associated with the $\Uc$ symmetry. This Lagrangian $\cL_{\cC}$ corresponds to a Chern--Simons theory with a new $K$-matrix 
\begin{align}
\tilde K=
\left(
\begin{array}{cc}
     0 & \bt^\sT\\
     \bt & K 
\end{array}
\right)
\end{align}
and the dynamical gauge field vector
\begin{align}
\tbb=
\left(
\begin{array}{c}
    \beta\\
    {\bf b} 
\end{array}
\right).
\end{align}
Note that $\det\tilde K=0$ because ${\bt}^{\sT}K^{-1}{\bt}=0$. 

To understand the implications of Eq. \eqref{eq:LagrangianCK}, we rewrite the gauge fields in a different basis. First, note that the null space of $\tilde K$ within the space of integer vectors, i.e., $
\left\{{\bf l}\in\mathbb{Z}^{N+1}\mid\tilde K{\bf l}=0\right\}$,
is generated by the integer vector
\begin{align}
\Lambda=
\left(
\begin{array}{c}
    -n\\
    nK^{-1}\bt 
\end{array}
\right),
\end{align}
where $n$ is the smallest positive integer for which $\Lambda$ is an integer vector. Physically, $n$ corresponds to the order of the vison $v\in\cal D$ because $n\bt$ can be expressed as an integer linear combination of the row vectors of $K$, indicating that the anyon with charge vector $n\bt$ under the gauge fields $\bb$ is a topologically trivial excitation in $\cD$. 

There exists an integer-valued matrix $W$ such that
\begin{align}
\det W=1,\qquad
W^{-1}\Lambda=
\left(
\begin{array}{c}
    1\\
    {\bf 0} 
\end{array}
\right)
\equiv\hat{\bf e}_0.
\end{align}
Note that $\det W=1$ guarantees that $W^{-1}$ is also an integer matrix, and the choice of $W$ is not unique. We use $W$ to define a new basis for the gauge fields,
\begin{align}
\tbb'=
\left(
\begin{array}{c}
    \beta'\\
    {\bf b}' 
\end{array}
\right)
\equiv W^{-1}\tbb.
\end{align}
The $K$-matrix in the new basis is $W^\sT\tilde K W=
\left(
\begin{array}{cc}
     0 & 0\\
     0 & \tilde{K}' 
\end{array}
\right),$
where the vanishing entries follow from $0
=W^\sT\tilde K W
\hat{\bf e}_0
=W^\sT\tilde K\Lambda$
and the fact that $W^\sT\tilde K W$ is symmetric. We can now rewrite the Lagrangian of $\cC$ in Eq. \eqref{eq:LagrangianCK} as
\begin{align}
  \cL_{\cC}[A]
  =\sum_{I,J=1}^N\frac{\tilde K'_{IJ}}{4\pi}b'_I db'_J
  -\frac{1}{2\pi}w_I Adb_I'
  +\frac{n}{2\pi}Ad\beta',
  \label{eq:LagrangianCK2}
\end{align}
where the integer coefficients $w_{I=1,2,\ldots,N}$ depend on the specific choice of $W$. Note that the gauge field $\beta'$ appears only in the coupling to the background gauge field $A$ associated with the global $\Uc$ symmetry. The coupling constant is fixed by the relation $n=-\hat{\bf e}_0^\sT W\hat{\bf e}_0
=-\hat{\bf e}_0^\sT\Lambda,$
which is independent of the specific choice of $W$. As a consequence of the last term in Eq. \eqref{eq:LagrangianCK2}, the $\Uc$ symmetry is spontaneously broken down to its $\mathbb{Z}_n^{(0)}$ subgroup in the TOSF $\cal C$. Once we add a Maxwell term for $\beta'$ to $\cL_{\cC}$, the photon fluctuations of $\beta'$ describe the Goldstone mode of the TOSF $\cal C$. Recall that $n$ is the order of the vison $v\in\cD$. Hence, the symmetry-breaking pattern of the TOSF $\cC$ obtained from this field-theoretic treatment is consistent with the algebraic approach in Sec. \ref{sec:GeneralFormalism}.

It is evident from Eq. \eqref{eq:LagrangianCK2} that the TOSF $\cal C$ contains a TO sector described by the Chern--Simons coupling of the gauge fields $\bb'$. The gapped anyons in this sector are particles that are charged under $\bb'$ but neutral under $\beta'$. In the original gauge-field basis $\tbb$, such a particle has a charge vector $\bf l$ satisfying $\Lambda^\sT{\bf l}=0$. The solutions for $\bf l$ take the general form
\begin{align}
{\bf l}=
\left(
\begin{array}{c}
    m\\
    {\bf x}   
\end{array}
\right),
\end{align}
where $\bf x$ is any $N$-component integer vector such that $m=\bt^\sT K^{-1}{\bf x}\in\mathbb{Z}.$
Physically, each such vector $\bf x$ corresponds precisely to an anyon in $\cD$ that has trivial braiding statistics, $e^{\ii 2\pi\bt^\sT K^{-1}{\bf x}}=1,$
with the vison $v$ before gauging $\Udual$ and hence remains deconfined after the condensation of $v$ in the gauging process. The braiding phase between two gapped anyons in $\cC$ with charge vectors ${\bf l}_{1,2}=
\left(
\begin{array}{c}
    m_{1,2}\\
    {\bf x}_{1,2}   
\end{array}
\right)$ is given by
\begin{align}
M_{\bl_1,\bl_2} = e^{\ii 2\pi{\bf l}_1^\sT\tilde K^{-1}\bl_2}
=e^{\ii 2\pi{\bf x}_1^\sT K^{-1}\bx_2},
\end{align}
which exactly matches that of the corresponding anyons with charge vectors ${\bf x}_{1,2}$ in $\cD$. Although $\tilde K$ is noninvertible, the quantity ${\bf l}_1^\sT\tilde K^{-1}\bl_2$ remains well defined because $\bl_{1,2}$ are both orthogonal to the null space of $\tilde K$. We therefore conclude that the gapped deconfined anyons in the TOSF $\cC$ are precisely described by the anyon condensate $\cal D/\cA$, where $\cA$ is the condensable algebra generated by the vison $v$. This result is consistent with Sec. \ref{sec:GeneralFormalism}.

The Lagrangian in Eq. \eqref{eq:LagrangianCK2} also indicates that the vortices of the TOSF $\cC$, which carry nonzero flux $dA/(2\pi)$, are charged under the gauge field $\beta'$. Objects charged under $\beta'$ have charge vectors that are not orthogonal to $\Lambda$ in the original gauge-field basis $\tbb$. They are not in the gapped deconfined anyons sector $\cD/\cA$. Instead, one can show that vortices with vorticities nonzero modulo $2\pi$ must descend from anyons in $\cD$ that braid nontrivially with the vison. In the present case, where the $\Udual$-enriched TO $\cD$ is Abelian, the vortices of $\cC$ do not carry non-Abelian degrees of freedom. Nevertheless, the field-theoretic approach here corroborates the algebraic analysis in Sec. \ref{sec:GeneralFormalism}.

\subsection{TOSF $\cC$ encoded by non-Abelian TO $\cD$}
\label{sec:nonAbelianD}

We first discuss examples of TOSFs encoded by non-Abelian TOs that exhibit non-Abelian vortices. We introduce these examples using parton constructions, which naturally yield their field-theoretic descriptions. We then show how these field theories complement the algebraic descriptions of these TOSFs. Finally, we provide a general field-theoretic formulation of TOSFs encoded by non-Abelian $\Udual$-enriched TOs.

\subsubsection{A charge-4 TOSF with non-Abelian vortices and its generalizations}
\label{sec:Charge4eExample}

Let us consider a concrete example of a charge-4 TOSF. Consider a 2D lattice of bosons with creation operator $\Psi^\dag({\bf r})$. The $\Uc$ symmetry of this system corresponds to particle-number conservation of $\Psi$. The filling of $\Psi$ is $n_\Psi=2$ bosons per site. To construct an ansatz for a TOSF $\cC$, we split $\Psi$ into two bosonic partons, $\varphi$ and $\chi$:
\begin{align}
    \Psi({\bf r})=\varphi({\bf r})\chi({\bf r}), 
\end{align}
where $\chi({\bf r})$ carries charge 1 under $\Uc$, while $\varphi({\bf r})$ is neutral. Both partons couple to a $\U(1)$ gauge field $b_0$, with the corresponding gauge transformation acting on the partons as $\varphi\rightarrow e^{\ii\theta}\varphi$ and $\chi\rightarrow e^{-\ii\theta}\chi$. Both $\varphi({\bf r})$ and $\chi({\bf r})$ have filling $n_\varphi=n_\chi=2$ per site on the lattice. 

We choose an ansatz in which the effective mean-field band structure for $\varphi$ is a Chern band with Chern number $C=1$. At filling $n_\varphi=2$, the bosons $\varphi$ can form the $\mathbb{Z}_4$ bosonic Read--Rezayi state\cite{ReadRezayi1999} (which can be obtained by symmetrizing four copies of the filling-$1/2$ bosonic Laughlin state\cite{Cappelli2001_para} or through a projective construction\cite{BarkeshliWen2010_projective}). The topological field theory description of this $\mathbb{Z}_4$ bosonic Read--Rezayi state is given by the non-Abelian $\U(2)$ Chern--Simons theory
\begin{align}
    \cL_\varphi
    =\frac{4}{4\pi}\Tr\left(\mfb d\mfb-\frac{2\ii}{3}\mfb^3\right) - \frac{4}{4\pi} (\Tr \mfb) \, (d\Tr\mfb),
    \label{eq:LagrangianVarphiCS}
\end{align}
with a dynamical $\U(2)$ gauge field $\mfb$:
\begin{align}
    \mfb
    =\frac{1}{2}b_0\mathds{1}
    +\frac{1}{2}b_1\sigma^x
    +\frac{1}{2}b_2\sigma^y
    +\frac{1}{2}b_3\sigma^z,
\end{align}
where $\sigma^{x,y,z}$ are the three Pauli matrices. Note that $b_0=\Tr\mfb$ is the $\U(1)$ gauge field, with standard normalization, that couples to both bosons $\varphi$ and $\chi$. Before allowing $b_0$ to fluctuate, the $\mathbb{Z}_4$ bosonic Read--Rezayi state of $\varphi$ exhibits the $\SU(2)_4$ TO. $\cL_{\varphi}$ effectively contains the term $-\frac{2}{4\pi}b_0db_0$, which encodes the Hall conductance $\sigma_{xy}^{\rm RR}=2$ of the $\mathbb{Z}_4$ bosonic Read--Rezayi state.

In our ansatz, the boson $\chi$ forms the elementary BIQH state with Hall conductance $-2$. Since $\chi$ is charged under both $\Uc$ and the gauge field $b_0=\Tr\mfb$, this BIQH state is described by the Lagrangian
\begin{align}
    \cL_{\chi}
    =\frac{2}{4\pi}
    \left(A - \Tr\mfb\right)
    \left(dA - d\Tr\mfb\right).
    \label{eq:LagrangianChi}
\end{align}

The TOSF $\cC$ is described by the total Lagrangian
\begin{align}
    \cL_\cC[A]=\cL_\varphi+\cL_\chi. 
\end{align}
Note that the self-Chern--Simons coupling of $b_0=\Tr\mfb$ cancels in $\cL_\cC[A]$. The field $\Tr\mfb$ only couples to the background field $A$ through $-\frac{2}{2\pi}A d\Tr\mfb$. Therefore, once a Maxwell term for $b_0$ is included in $\cL_\cC[A]$, the dynamics of $b_0=\Tr\mfb$ give rise to gapless photons, which provide a dual description of the Goldstone mode of the TOSF $\cal C$ with spontaneously broken $\Uc$ symmetry. 

Determining the unbroken subgroup of $\Uc$ requires some care. Naively, the coupling $-\frac{2}{2\pi}A d\Tr\mfb$ appears to suggest that the unbroken subgroup is $\mathbb{Z}_2$. However, a configuration with an odd flux quantum of $\Tr\mfb$ can be identified with a gauge flux in the $\SU(2)$ subgroup of the $\U(2)$ gauge field $\mfb$, which remains gapped by the Chern--Simons term in $\cL_\varphi$. Hence, the gapless photon fluctuations of $b_0=\Tr\mfb$ break $\Uc$ only down to its $\mathbb{Z}_4$ subgroup. 

The topological sector of gapped anyons in $\cC$, with a trivial $\Uc$ background $A=0$ and zero vorticity, must be neutral under the gapless gauge field $b_0$. Hence, this sector consists only of the integer-spin representations of the $\SU(2)$ part of the gauge group, namely, those with spins $j=0,1,2$. Here, $j$ cannot exceed 2 because the $\SU(2)$ Chern--Simons level in Eq. \eqref{eq:LagrangianVarphiCS} is 4. Thus, the topological sector is $\SU(2)_4/\mathbb{Z}_2=\SO(3)_4=\SU(3)_1$. The two nontrivial Abelian anyons in $\SU(3)_1$ arise from the splitting of the non-Abelian anyon with $j=1$.

The elementary vortex of $\cC$ carries a flux $2\pi/4$ of $A$. The coupling $-\frac{2}{2\pi}A d\Tr\mfb$ implies that the elementary vortex carries charge $-1/2$ under $\Tr\mfb=b_0$ and can therefore be identified with either the $j=1/2$ or $j=3/2$ representation of the $\SU(2)$ gauge group. These representations correspond to anyons with quantum dimension $\sqrt{3}$ in the $\SU(2)_4$ TO. The elementary vortex of $\cC$ is expected to inherit this quantum dimension. Phenomenologically, this TOSF resembles the charge-$4e$ topological superconductor in a fermionic system recently discussed in Ref. \onlinecite{ShiHan4e2026}. However, our parton construction differs substantially from the systems studied there.

Having established the parton construction and field theory for this TOSF $\cC$, we now discuss how the topological data of $\cC$ is encoded by a $\Udual$-enriched TO $\cD$. Gauging $\Uc$ to obtain $\cD$ amounts to promoting $A$ to a dynamical $\U(1)$ gauge field $\alpha$, resulting in the Lagrangian
\begin{align}
    \cL_\cD
    ={}&\frac{4}{4\pi}
    \Tr\left(\mfb d\mfb-\frac{2\ii}{3}\mfb^3\right) - \frac{4}{4\pi} \Tr \mfb \, (d\Tr\mfb)
    \nonumber\\
    &+\frac{2}{4\pi}
    \left(\alpha- \Tr\mfb\right)
    \left(d\alpha- d\Tr\mfb\right)
    +\frac{1}{2\pi}B d\alpha.
\end{align}
By shifting $\alpha\rightarrow\alpha+\Tr\mfb$, we find that the TO $\cD$ is given by
\begin{align}
    \cD=\U(2)_{4,-4}\boxtimes\U(1)_{2} =\frac{\SU(2)_4\boxtimes\U(1)_{-8}}{\mathbb{Z}_2}
    \boxtimes\U(1)_{2}.
\end{align}
The convention for the Chern-Simons levels of $\U(2)_{4,-4}$ and the rewriting in the second equality follow Ref.~\onlinecite{Hsin2016LevelRankDuality}.
Using this representation, the anyons in $\cD$ can be labeled by triplets $(j,q,s)$. Here, $j=0,\frac{1}{2},1,\frac{3}{2},2$ labels the spin of the $\SU(2)$ representation in the $\SU(2)_4$ sector, while $q=0,1,\ldots,7~({\rm mod}~8)$ labels the charge in the $\U(1)_{-8}$ sector. The $\mathbb{Z}_2$ quotient imposes the constraint $2j=q\pmod{2}$ and the identification $(j,q,s) \sim (2-j,q+4,s) $. Finally, $s=0,1~({\rm mod}~2)$ labels the charge in the $\U(1)_{2}$ sector. The vison $v\in\cD$ is
$v=(j=0,q=2,s=1)$, which has order $n=4$ and bosonic self-statistics. 

According to Sec. \ref{sec:GeneralFormalism}, the gapped anyons in the TOSF $\cC$ are described by $\cD/\cA$, where $\cA$ is the condensable algebra generated by the vison $v$. In this case, $\cD/\cA=\SO(3)_4=\SU(3)_1$. The TOSF $\cC$ spontaneously breaks $\Uc$ down to $\mathbb{Z}_{n=4}^{(0)}$. Its elementary vortex is identified with the $\mathbb{Z}_4^{(0)}$ symmetry flux of $\cD/\cA$. This symmetry flux descends from the anyons
$\left(
    j=\frac{1}{2}\ {\rm or}\ \frac{3}{2},
    q\in2\mathbb{Z}+1,
    s \in \{0,1\}
    \right)\in\cD,$
which are confined by the condensation of the vison $v$ and have quantum dimension $\sqrt{3}$. These results are consistent with the field-theoretic analysis above.

This construction generalizes to TOSFs with higher-charge condensates. For example, at filling $n_\Psi=2$, one can consider an ansatz in which the bosonic parton $\varphi$ forms the bosonic non-Abelian spin-singlet (NASS) state with $\SU(3)_3$ TO\cite{Ardonne1999,Ardonne2001}, while the bosonic parton $\chi$ forms the BIQH state with Hall conductance $-2$. The resulting TOSF $\cC$ spontaneously breaks $\Uc$ down to a $\mathbb{Z}_6^{(0)}$ subgroup, contains a gapped TO sector
$\SU(3)_3/\mathbb{Z}_3={\rm Spin}(8)_1,$
namely, the TO of the so-called ``three-fermion toric code,'' and supports vortices with quantum dimension $2$. Within this construction, one can generalize the state of $\varphi$ to more elaborate NASS states to obtain TOSFs with more exotic vortices.

\subsubsection{An alternative construction by doping a gapped quantum charge liquid}

We now discuss an alternative construction of the TOSF studied in Sec. \ref{sec:Charge4eExample} by doping a gapped topological quantum charge liquid.

Consider a bosonic system on a 2D lattice with two
orbitals per unit cell, labeled by $l=1,2$. There is no orbital symmetry in the system. The boson creation operators of the two orbitals are denoted by $\Psi^\dag_l({\bf r})$. The total filling is $n_{\Psi_1} + n_{\Psi_2}  = 2-\delta$ bosons per unit cell. We use the parton decomposition
\begin{align}
    \Psi_l(\mathbf r) = 
    \varphi_l(\mathbf r)h^\dagger(\mathbf r),
\end{align}
where the bosonic parton $h({\bf r})$ carries charge $-1$ under $\Uc$ and is interpreted as a holon, while the bosonic partons $\varphi_l({\bf r})$ are neutral under $\Uc$. All partons couple to a $\U(1)$ gauge field $b_0$, with the corresponding gauge transformation acting as $\varphi_l\rightarrow e^{\ii\theta}\varphi_l,~
    h\rightarrow e^{\ii\theta}h$. The gauge field $b_0$ enforces the local constraint $n_h +n_{\varphi_1} +n_{\varphi_2} =2$ on the parton filling per unit cell. Consequently, the average fillings satisfy
\begin{align}
    n_h=\delta,
    \qquad
   n_{\varphi_1} +n_{\varphi_2} =2-\delta.
\end{align}

At zero doping, $\delta=0$, the total filling of the neutral partons is $n_{\varphi_1}+n_{\varphi_2}=2$ per unit cell. We choose an orbital-mixing ansatz in which the single-particle states of $\varphi_1$ and $\varphi_2$ hybridize to form a Chern band with Chern number $C=1$. The neutral bosons occupy this band and can form the filling-$2$ bosonic $\mathbb{Z}_4$ Read--Rezayi state. This state has Hall conductance $\sigma_{xy}^{\rm RR}=2$ with respect to the $\U(1)$ gauge field $b_0$. The holons are at zero density and therefore produce a trivial response to $A$ or $b_0$. This ansatz yields a gapped topological quantum charge liquid of $\Psi_l$ with a trivial response under $\Uc$ and nontrivial $\U(2)_{4,-4}$ TO. The $\U(2)_{4,-4}$ TO arises when $b_0$ is made dynamical in the effective theory of the $\mathbb{Z}_4$ Read--Rezayi state, whose Lagrangian is given by Eq. \eqref{eq:LagrangianVarphiCS}.

We now dope this quantum charge liquid by taking $\delta>0$ to be small but finite. Define the normalized $b_0$ flux per unit cell by $ \Phi_{b_0} \equiv\frac{1}{2\pi}\int_{\rm u.c.}db_0.$
We assume that the bosons $\varphi_l$ remain in the $\mathbb{Z}_4$ Read--Rezayi state in the presence of a uniform additional flux
\begin{align}
    \Phi_{b_0}=-\frac{\delta}{2}.
\end{align}
Indeed, the corresponding flux--density relation gives
\begin{align}
    n_{\varphi_1} +n_{\varphi_2}-2
    =\sigma_{xy}^{\rm RR}\Phi_{b_0}
    =2\left(-\frac{\delta}{2}\right)
    =-\delta.
\end{align}
The holons have filling $n_h=\delta$ and experience the same flux $\Phi_{b_0}=-\delta/2$. They can therefore form the BIQH state with Hall conductance $-2$.
The corresponding Lagrangian is
\begin{align}
    \cL_h
    =\frac{2}{4\pi}
    \left(-A+\Tr\mfb\right)
    \left(-dA+d\Tr\mfb\right).
\end{align}
This Lagrangian is identical to Eq. \eqref{eq:LagrangianChi}. Therefore, an analysis parallel to that in Sec. \ref{sec:Charge4eExample} shows that, within this parton construction, doping the topological quantum charge liquid produces the same charge-4 TOSF phase with non-Abelian vortices for $\delta>0$.

\subsubsection{A general field-theoretic description and a natural phase transition}

We now develop a field-theoretic description of the TOSF $\cC$ by carefully gauging $\Udual$ in the field theory of the TO $\cD$. The resulting theory corroborates the algebraic approach in Sec.~\ref{sec:GeneralFormalism} and suggests a natural quantum phase transition out of the TOSF phase.

As discussed above, the $\Udual$ enrichment of the TO $\cD$ is characterized by the vison $v$, namely, the Abelian anyon generated by a $2\pi$ flux of the $\Udual$ background gauge field. If the vison $v\in\cD$ has order $n$, it generates a $\mathbb{Z}_n^{(1)}$ 1-form symmetry. We denote the associated $\mathbb{Z}_n$-valued 2-form background gauge field by $V^{(2)}$. The coupling of $\cD$ to the $\Udual$ background gauge field $B$ can be more precisely expressed as $\cL_{\cD}[V^{(2)}]$, namely the Lagrangian of $\cD$ with the coupling to the background $V^{(2)}$, subject to the constraint 
\begin{align}
    V^{(2)}=\frac{1}{2\pi}dB\pmod{n}.
    \label{eq:U1enrichment}
\end{align}
Here, the equality is understood at the level of cocycles. Equivalently, we can use the following Lagrangian for $\cD$:
\begin{align}
    \cL_{\cD}'[B]
    =\cL_{\cD}[\nu^{(2)}]
    +\frac{2\pi}{n}\omega^{(1)}
    \cup\left(\nu^{(2)}-\frac{dB}{2\pi}\right),
    \label{eq:LagrangianDprime}
\end{align}
where $\omega^{(1)}$ and $\nu^{(2)}$ are dynamical $\mathbb{Z}_n$-valued 1- and 2-form gauge fields, respectively, represented as cocycles, and $\cup$ denotes the cup product. Integrating out $\omega^{(1)}$ identifies $\nu^{(2)}$ with the background field $V^{(2)}$ satisfying Eq. \eqref{eq:U1enrichment}. For $\cL_{\cD}'$ to be well defined for arbitrary configurations of $\omega^{(1)}$ and $\nu^{(2)}$, the $\mathbb{Z}_n^{(1)}$ 1-form symmetry of $\cD$ must be nonanomalous. Equivalently, the vison must have bosonic self-statistics, which is indeed a prerequisite for a $\Udual$-enriched TO $\cD$ to encode the topological data of a TOSF through gauging.

After gauging the $\Udual$ symmetry, we obtain the following Lagrangian for the TOSF $\cC$:
\begin{align}
    \cL_{\cC}[A]
    ={}&\cL_{\cD}[\nu^{(2)}]
    +\frac{2\pi}{n}\omega^{(1)}
    \cup\left(\nu^{(2)}-\frac{d\beta}{2\pi}\right)
    -\frac{1}{2\pi}A d\beta,
\end{align}
where a Maxwell term for $\beta$ is implicitly included. Since $\beta$ has no Chern--Simons self-coupling and couples only to the discrete gauge field $\omega^{(1)}$ and the background field $A$, its fluctuations give rise to gapless photons. Using particle--vortex duality,\cite{Dasgupta1981_latticePV,Peskin1978_latticePV} the TOSF $\cC$ can equivalently be described by
\begin{align}
    \cL_{\cC}'[A]
    ={}&\cL_{\cD}[\nu^{(2)}]
    +\frac{2\pi}{n}\omega^{(1)}\cup\nu^{(2)}
    \nonumber\\
    &+\left|D_{\frac{2\pi}{n}\omega^{(1)}+A}\Phi\right|^2
    -\mu|\Phi|^2+\cdots,
    \label{eq:LagrangianCprime}
\end{align}
where $\Phi$ is a complex scalar field whose conserved current is dual to $d\beta/(2\pi)$. The $\Uc$ symmetry corresponds to conservation of the particle number of $\Phi$. The field $\Phi$ carries unit charge under both the $\Uc$ background gauge field $A$ and the dynamical $\mathbb{Z}_n$ gauge field $\omega^{(1)}$. The term $\left|D_{\frac{2\pi}{n}\omega^{(1)}+A}\Phi\right|^2$ denotes the kinetic term of $\Phi$. We take $\mu<0$, so that $\Phi$ is condensed, with $\langle\Phi\rangle\neq0$. Recall that the Lagrangian is written for real time. The Goldstone mode of this condensate is the Goldstone mode of the TOSF and is dual to the gapless photon of $\beta$. The ellipsis denotes higher-order terms in $|\Phi|^2$.

Because $\Phi$ couples to both $A$ and the $\mathbb{Z}_n$ gauge field $\omega^{(1)}$, its condensation spontaneously breaks $\Uc$ only down to its $\mathbb{Z}_n^{(0)}$ subgroup, which is captured by the local gauge-invariant order parameter $\langle\Phi^n\rangle\neq 0$. This condensation higgses $\omega^{(1)}$, leaving the $\mathbb{Z}_n$ 2-form gauge field $\nu^{(2)}$ unconstrained and free to fluctuate. Consequently, $\cL_{\cD}[\nu^{(2)}]$ describes a gapped sector of the TOSF $\cC$ with TO $\cD/\cA$, namely $\cD$ with its $\mathbb{Z}_n^{(1)}$ 1-form symmetry gauged. This TO is enriched by the unbroken $\mathbb{Z}_n^{(0)}$ 0-form symmetry whose background gauge field is effectively represented by $\omega^{(1)}$. The corresponding $\mathbb{Z}_n^{(0)}$ symmetry flux is identified with a vortex in the TOSF carrying $2\pi/n$ flux of $\Uc$ as a consequence of the condensation of $\Phi$.

The field theory in Eq. \eqref{eq:LagrangianCprime} also suggests a natural phase transition out of the TOSF $\cC$. As $\mu$ changes sign from negative to positive, $\Phi$ transitions from the condensed phase to a gapped phase with $\langle\Phi\rangle=0$. When $\Phi$ is gapped, $\omega^{(1)}$ fluctuates freely, and consequently constrains the $\mathbb{Z}_n$ 2-form gauge field to $\nu^{(2)}=0$. The resulting phase is therefore a gapped phase topologically equivalent to the TO $\cD$. This phase transition is an example of the Abelian-anyon proliferation transition discussed in Ref. \onlinecite{ChengSeiberg2026}.

\section{Summary and Outlook}
\label{sec:conclusion}

In this work, we develop a general framework for characterizing the topological data of $(2+1)$D TOSFs. The framework is based on a two-way mapping, implemented by gauging the respective $\U(1)$ symmetries, between TOSFs with spontaneously broken global $\Uc$ symmetry and $\Udual$-enriched TOs with vanishing Hall conductance. Through this mapping, the topological data of a TOSF $\cC$, including the charge of the local condensed object, the deconfined anyon content, and the topological properties of vortices, can be systematically characterized in terms of the corresponding $\Udual$-enriched TO $\cD$. We develop both an algebraic formulation and a complementary field-theoretic description of this framework. One interesting result of this framework is the identification of the topological properties of TOSF vortices with those of discrete symmetry fluxes in the TOSF's gapped anyon sector treated as a TO enriched the residual discrete symmetry. This identification is consistent with the argument based on continuously parametrized families of TOs in Ref.~\onlinecite{Else2021TopoGoldstone} and provides a unified perspective on topological quantum computation using discrete symmetry defects in gapped TOs and vortices in TOSFs in bosonic systems. In addition, our field-theoretic formulation further suggests a natural phase transition from $\cC$ to a gapped phase with the same intrinsic TO as $\cD$. Finally, using parton constructions and field-theoretic approaches, we provide new examples of TOSFs with exotic non-Abelian vortices.

Our results suggest several directions for future investigation. One important direction is to identify microscopic models that realize TOSFs with non-Abelian vortices. Although we construct such states using partons, identifying microscopic Hamiltonians that energetically stabilize these phases remains an open question. Such models would deepen our understanding of topological phases beyond fully gapped systems and could provide new platforms for topological quantum computation.

A natural extension of our framework is to fermionic topologically ordered superconductors with spontaneously broken charge $\Uc$ symmetry. It is well known that a fermionic $p_x+ip_y$-wave superconductor has no intrinsic deconfined anyons (before gauging), yet its elementary vortices bind Majorana zero modes and exhibit non-Abelian braiding\cite{ReadGreen2000,Ivanov2001}. In the presence of deconfined anyons, superconducting vortices can exhibit even richer topological properties, as illustrated by the examples in Refs.~\onlinecite{ShiHan4e2026,GaoWu2025,Gao2026}. Conceptually, gauging the charge $\Uc$ symmetry should again eliminate the Goldstone mode through the Higgs mechanism and map these superconductors to gapped TOs. Implementing this mapping, however, requires extending the gauging framework of Ref.~\onlinecite{GaugingU1}, which was developed for bosonic systems, to accommodate local fermionic excitations. Such an extension would allow us to investigate how the resulting TO encodes both the deconfined anyon content and the topological properties of vortices in the original fermionic superconductor.

Another direction is to extend the investigation to topologically ordered gapless phases in which deconfined, gapped anyons coexist with spontaneous breaking of a general continuous global symmetry $G$. A central challenge is to systematically characterize the topological properties of order-parameter defects (and/or textures) and their interplay with deconfined anyons. In a TOSF with spontaneously broken $\Uc$ symmetry, the topological defects are vortices labeled by integer vorticities. More generally, when the continuous symmetry $G$ is spontaneously broken to a subgroup $H$, the order-parameter manifold is $G/H$, whose homotopy groups $\pi_k(G/H)$ provide a starting point for characterizing the possible defects.

A familiar example is provided by quantum Hall ferromagnets in the spin-rotation-invariant limit~\cite{Sondhi1993}. Here, $G=\U(2)$ combines spin rotations and charge conservation, while the unbroken subgroup $H=\U(1)_{\uparrow}\times\U(1)_{\downarrow}$ corresponds to independent conservation of the numbers of electrons with spins along and against the ferromagnetic polarization direction. The order-parameter manifold is $G/H = S^2$, which supports skyrmion textures classified by $\pi_2(S^2)=\mathbb{Z}$. In fractional quantum Hall ferromagnets, these skyrmions can carry fractional electric charge, reflecting their interplay with the underlying TO. A natural extension is to consider phases with uniaxial spin-quadrupolar order, for which the unbroken subgroup is $H=(\U(1)_{\uparrow}\times\U(1)_{\downarrow})\rtimes\mathbb{Z}_2$. The additional $\mathbb{Z}_2$ symmetry interchanges the two spin species, yielding the order-parameter manifold $G/H = \mathbb{RP}^2$. This manifold supports both $\mathbb{Z}_2$ vortex defects associated with $\pi_1(\mathbb{RP}^2)=\mathbb{Z}_2$ and skyrmion textures associated with $\pi_2(\mathbb{RP}^2)=\mathbb{Z}$. Systematically characterizing the topological data of phases in which deconfined anyons coexist with such order thus constitutes a concrete instance of this broader problem of topologically ordered phases with spontaneous continuous symmetry breaking.

A potential general approach is again to map the gapless symmetry-broken phase $\cC$ to a gapped TO $\cD$ by gauging $G$, as in our treatment of TOSFs. A framework for gauging an unbroken continuous Lie group symmetry in a $(2+1)$D TO was developed in Ref.~\onlinecite{GaugingLieGroup}. Extending this framework to phases in which $G$ is spontaneously broken to $H$ could provide such a mapping, enabling us to investigate how the topological data of $\cD$ encode the topological properties of order-parameter defects and textures in $\cC$, including their interplay with the deconfined anyons.

\section*{Acknowledgement}
We thank Meng Cheng and Zhaoyu Han for helpful discussions. This work is supported in part by the Alfred P. Sloan Foundation through a Sloan Research Fellowship and by the National Science Foundation under Award No.~PHY-2609907. We acknowledge the use of ChatGPT (OpenAI) for language polishing and proofreading of this manuscript. We reviewed and edited all generated text and take full responsibility for the final content.

{\it Note added} - While preparing this manuscript, we became aware of independent, related works~\cite{ChengSeiberg2026Peeling,HanTOSF}. Ref. \onlinecite{HanTOSF} is expected to appear on arXiv on the same day as ours. We thank the authors for informing us of their work in advance.

\appendix
\section{Different $\Udual$-enriched TOs encoding the same TOSF $\cC$}
\label{app:Dprime}

In Sec. \ref{sec:GeneralFormalism}, we assume that stacking the TOSF $\cC$ with a $\Uc$-symmetric BIQH state does not change its topological class. However, such stacking changes the $\Udual$-enriched TO obtained by gauging $\Uc$. Below, we show that this change is given by Eq. \eqref{eq:Dprime}.

Upon stacking $\cC$ with a BIQH state of Hall conductance $p\neq 0$ and gauging $\Uc$, we obtain a $\Udual$-enriched phase $\cD'$ with Lagrangian
\begin{align}
    \cL_{\cD'}[B]
    =\cL_{\cal C}[\alpha]
    +\frac{-p}{4\pi}\alpha d\alpha
    +\frac{1}{2\pi}B d\alpha,
\end{align}
where the Chern--Simons term $\frac{-p}{4\pi}\alpha d\alpha$ describes the contribution from the BIQH state. An equivalent description is
\begin{align}
    \cL_{\cD'}'[B]
    ={}&\cL_{\cal C}[\alpha]
    +\frac{-p}{4\pi}\alpha' d\alpha'
    +\frac{1}{2\pi}\gamma(d\alpha-d\alpha')
    +\frac{1}{2\pi}B d\alpha,
\end{align}
where $\alpha'$ and $\gamma$ are dynamical $\U(1)$ gauge fields. Integrating out $\gamma$ imposes $\alpha=\alpha'$ and recovers $\cL_{\cD'}[B]$.

The Lagrangian $\cL_{\cD'}'[B]$ can also be interpreted as follows. We first gauge $\Uc$ in $\cC$ and in the BIQH state separately, obtaining $\cD$ and a $\U(1)_{-p}$ TO, respectively, each with its own dual $\U(1)$ symmetry. The diagonal subgroup of these two dual symmetries is identified with the eventual $\Udual$ symmetry of $\cD'$, while the relative subgroup, denoted by $\Ur$ and associated with the gauge field $\gamma$, is gauged. In other words, $\cD'$ is obtained by gauging the relative $\Ur$ symmetry of $\cD\boxtimes\U(1)_{-p}$.

Before gauging $\Ur$, we first characterize how it enriches $\cD\boxtimes\U(1)_{-p}$. We label the anyons in $\cD\boxtimes\U(1)_{-p}$ by $ (x,y)$, where $x\in \cD$ denotes an anyon type in the $\cD$ sector, while $y=0,1,\ldots,|p|-1 \mod |p|$ labels an anyon type in the Abelian $\U(1)_{-p}$ sector. The vison associated with $\Ur$ is
\begin{align}
    w=(v,-1),
\end{align}
whose order is
\begin{align}
    n_w= n|p|/l,
\end{align}
where $l=\gcd(n,|p|)$. The TO $\cD\boxtimes\U(1)_{-p}$ has Hall conductance
\begin{align}
    \sigma_{xy}^{\rm r}=-1/p
\end{align}
with respect to $\Ur$. This Hall conductance arises entirely from the $\U(1)_{-p}$ sector because $\cD$ has vanishing Hall conductance with respect to its own $\Udual$ symmetry.

Following the general framework for gauging $\U(1)$ symmetries in 2+1D topological orders,\cite{GaugingU1} gauging $\Ur$ gives
\begin{align}
    \cD'
    =\frac{
    \cD\boxtimes\U(1)_{-p}
    \boxtimes\U(1)_{-\sigma_{xy}^{\rm r}n_w^2}
    }{
    \langle (v,-1,n_w|\sigma_{xy}^{\rm r}|)\rangle
    }.
    \label{eq:DprimeApp}
\end{align}
Note that we use a slightly different sign convention for the Hall conductance and the Chern--Simons level than in Ref. \onlinecite{GaugingU1}.
Let us unpack this expression [Eq. \eqref{eq:DprimeApp}]. The numerator is $\cD\boxtimes\U(1)_{-p}$ augmented by an additional Abelian TO sector $\U(1)_{-\sigma_{xy}^{\rm r}n_w^2} =\U(1)_{n^2p/l^2}$,
which arises from the nonzero Hall conductance $\sigma_{xy}^{\rm r}$ with respect to $\Ur$. An anyon in $ \cD\boxtimes\U(1)_{-p}\boxtimes\U(1)_{n^2p/l^2}$
can be labeled by a triplet $(x,y,z)$, where $x\in\cD$, $y=0,1,\ldots,|p|-1 ~~({\rm mod}~|p|)$, and $z=0,1,\ldots,n^2|p|/l^2-1 ~~({\rm mod}~n^2|p|/l^2)$. $y$ and $z$ label the anyons in the $\U(1)_{-p}$ and $\U(1)_{n^2p/l^2}$ sectors, respectively. The anyon appearing in the denominator of Eq. \eqref{eq:DprimeApp} is $ (v,-1,n_w|\sigma_{xy}^{\rm r}|)=\left(v,-1,n/l\right)$, which is an Abelian anyon with bosonic self-statistics. The bracket $\langle \cdot \rangle$ represents the condensable algebra generated by the argument. 
The quotient in Eq. \eqref{eq:DprimeApp} denotes the anyon condensation as before. 
Eq. \eqref{eq:DprimeApp} therefore agrees with Eq. \eqref{eq:Dprime} for the $\Udual$-enriched TO $\cD'$. The $\Udual$ vison $v'$ in $\cD'$ can be represented by the anyon $ v'=(v,0,0)
    \in\cD\boxtimes\U(1)_{-p}
    \boxtimes\U(1)_{n^2p/l^2}$.
The vison $v'$ has order $n$, the same as the original vison $v\in\cD$.

As a sanity check, one can verify that gauging the $\Udual$ symmetry of $\cD'$ recovers the TOSF $\cC$. Like $\cD$, the TO $\cD'$ has vanishing Hall conductance with respect to $\Udual$. Since the vison $v'$ has order $n$, gauging $\Udual$ therefore spontaneously breaks the $\Uc$ symmetry down to its $\mathbb{Z}_n^{(0)}$ subgroup. The gapped topological sector of $\cC$ is recovered by condensing the vison $v'$ in $\cD'$. This condensation amounts to condensing the algebra generated by both $(v,0,0)$ and $\left(v,-1,n/l\right)$ in $\cD\boxtimes\U(1)_{-p}\boxtimes\U(1)_{n^2p/l^2}$. The resulting topological order is $\cD/\cA$. The $\U(1)_{-p}\boxtimes\U(1)_{n^2p/l^2}$ sector is completely removed by the condensation. The elementary $2\pi/n$ vortex of the TOSF $\cC$ is again identified with the $\mathbb{Z}_n^{(0)}$ symmetry flux of $\cD/\cA$.

Lastly, we examine the dependence of $\cD'$ on $p$. $\cD'$ is obtained by condensing the algebra generated by $\left(v,-1,n/l\right)$ in $\cD\boxtimes\U(1)_{-p}\boxtimes\U(1)_{n^2p/l^2}$. Since $\left(v,-1,n/l\right)^n=\left({\rm id},-n,n^2/l\right)$, this anyon and its powers form a subalgebra of the condensate supported entirely in the $\U(1)_{-p}\boxtimes\U(1)_{n^2p/l^2}$ sector. Condensing this subalgebra reduces the two Abelian factors to the anyon condensate ${\cal B}_{n,p}\equiv\frac{\U(1)_{-p}\boxtimes\U(1)_{n^2p/l^2}}{\langle(-n,n^2/l)\rangle}$, which turns out to exhibit the Abelian TO of a twisted $\mathbb{Z}_n$ gauge theory in 2+1D. ${\cal B}_{n,p}$ is generated by two deconfined anyons $h_1=(-1,n/l)$ and $h_2=(0,p/l)$, where these labels denote representatives in $\U(1)_p\boxtimes\U(1)_{-n^2p/l^2}$. Every anyon in ${\cal B}_{n,p}$ can be expressed as $h_1^{m_1} h_2^{m_2}$, subject to the identifications $h_1^n\sim(0,0)$ and $h_2^n\sim h_1^p$. Their braiding data are determined by $M_{h_1,h_2}=e^{\ii2\pi/n}$, $\theta_{h_1}=1$, and $\theta_{h_2}=e^{\ii\pi p/n^2}$. Here, $M_{h_1,h_2}$ is the braiding statistics between $h_1$ and $h_2$. $\theta_{h_{1,2}}$ is the topological spin of $h_{1,2}$. These data show that, for fixed $n$, ${\cal B}_{n,p}$ depends only on $p\mod 2n$: the shift $p\rightarrow p+2n$ can be absorbed by the redefinition $h_2\rightarrow h_1h_2$. 
% Equivalently, the twist class is labeled by $p/2\mod n$. 
The anyon $h_1\in{\cal B}_{n,p}$ remains fixed under this shift. We can therefore rewrite $\cD'$ as
\begin{align}
    \cD'
    =\frac{\cD\boxtimes{\cal B}_{n,p}}
    {\langle(v,h_1)\rangle}.
\end{align}
This form makes manifest that, for fixed $\cD$ and $n$, the $\Udual$-enriched TO $\cD'$ depends only on $p\mod 2n$.

\bibliography{ref}

\begin{thebibliography}{56}%
\makeatletter
\providecommand \@ifxundefined [1]{%
 \@ifx{#1\undefined}
}%
\providecommand \@ifnum [1]{%
 \ifnum #1\expandafter \@firstoftwo
 \else \expandafter \@secondoftwo
 \fi
}%
\providecommand \@ifx [1]{%
 \ifx #1\expandafter \@firstoftwo
 \else \expandafter \@secondoftwo
 \fi
}%
\providecommand \natexlab [1]{#1}%
\providecommand \enquote  [1]{``#1''}%
\providecommand \bibnamefont  [1]{#1}%
\providecommand \bibfnamefont [1]{#1}%
\providecommand \citenamefont [1]{#1}%
\providecommand \href@noop [0]{\@secondoftwo}%
\providecommand \href [0]{\begingroup \@sanitize@url \@href}%
\providecommand \@href[1]{\@@startlink{#1}\@@href}%
\providecommand \@@href[1]{\endgroup#1\@@endlink}%
\providecommand \@sanitize@url [0]{\catcode `\\12\catcode `\$12\catcode `\&12\catcode `\#12\catcode `\^12\catcode `\_12\catcode `\%12\relax}%
\providecommand \@@startlink[1]{}%
\providecommand \@@endlink[0]{}%
\providecommand \url  [0]{\begingroup\@sanitize@url \@url }%
\providecommand \@url [1]{\endgroup\@href {#1}{\urlprefix }}%
\providecommand \urlprefix  [0]{URL }%
\providecommand \Eprint [0]{\href }%
\providecommand \doibase [0]{http://dx.doi.org/}%
\providecommand \selectlanguage [0]{\@gobble}%
\providecommand \bibinfo  [0]{\@secondoftwo}%
\providecommand \bibfield  [0]{\@secondoftwo}%
\providecommand \translation [1]{[#1]}%
\providecommand \BibitemOpen [0]{}%
\providecommand \bibitemStop [0]{}%
\providecommand \bibitemNoStop [0]{.\EOS\space}%
\providecommand \EOS [0]{\spacefactor3000\relax}%
\providecommand \BibitemShut  [1]{\csname bibitem#1\endcsname}%
\let\auto@bib@innerbib\@empty
%</preamble>
\bibitem [{\citenamefont {Kitaev}(2006)}]{Kitaev2006}%
  \BibitemOpen
  \bibfield  {author} {\bibinfo {author} {\bibfnamefont {Alexei}\ \bibnamefont {Kitaev}},\ }\bibfield  {title} {\enquote {\bibinfo {title} {Anyons in an exactly solved model and beyond},}\ }\href {\doibase 10.1016/j.aop.2005.10.005} {\bibfield  {journal} {\bibinfo  {journal} {Annals of Physics}\ }\textbf {\bibinfo {volume} {321}},\ \bibinfo {pages} {2–111} (\bibinfo {year} {2006})}\BibitemShut {NoStop}%
\bibitem [{\citenamefont {Etingof}\ \emph {et~al.}(2010)\citenamefont {Etingof}, \citenamefont {Nikshych}, \citenamefont {Ostrik},\ and\ \citenamefont {Meir}}]{EtingofNOM2009}%
  \BibitemOpen
  \bibfield  {author} {\bibinfo {author} {\bibfnamefont {Pavel}\ \bibnamefont {Etingof}}, \bibinfo {author} {\bibfnamefont {Dmitri}\ \bibnamefont {Nikshych}}, \bibinfo {author} {\bibfnamefont {Victor}\ \bibnamefont {Ostrik}}, \ and\ \bibinfo {author} {\bibfnamefont {Ehud}\ \bibnamefont {Meir}},\ }\bibfield  {title} {\enquote {\bibinfo {title} {{Fusion Categories and Homotopy Theory}},}\ }\href {\doibase 10.4171/QT/6} {\bibfield  {journal} {\bibinfo  {journal} {Quantum Topology}\ }\textbf {\bibinfo {volume} {1}},\ \bibinfo {pages} {209--273} (\bibinfo {year} {2010})},\ \Eprint {http://arxiv.org/abs/0909.3140} {arXiv:0909.3140 [math.QA]} \BibitemShut {NoStop}%
\bibitem [{\citenamefont {Barkeshli}\ \emph {et~al.}(2019)\citenamefont {Barkeshli}, \citenamefont {Bonderson}, \citenamefont {Cheng},\ and\ \citenamefont {Wang}}]{SET}%
  \BibitemOpen
  \bibfield  {author} {\bibinfo {author} {\bibfnamefont {Maissam}\ \bibnamefont {Barkeshli}}, \bibinfo {author} {\bibfnamefont {Parsa}\ \bibnamefont {Bonderson}}, \bibinfo {author} {\bibfnamefont {Meng}\ \bibnamefont {Cheng}}, \ and\ \bibinfo {author} {\bibfnamefont {Zhenghan}\ \bibnamefont {Wang}},\ }\bibfield  {title} {\enquote {\bibinfo {title} {Symmetry fractionalization, defects, and gauging of topological phases},}\ }\href {\doibase 10.1103/PhysRevB.100.115147} {\bibfield  {journal} {\bibinfo  {journal} {Phys. Rev. B}\ }\textbf {\bibinfo {volume} {100}},\ \bibinfo {pages} {115147} (\bibinfo {year} {2019})},\ \Eprint {http://arxiv.org/abs/arXiv:1410.4540} {arXiv:1410.4540} \BibitemShut {NoStop}%
\bibitem [{\citenamefont {Teo}\ \emph {et~al.}(2015)\citenamefont {Teo}, \citenamefont {Hughes},\ and\ \citenamefont {Fradkin}}]{TeoSET2015}%
  \BibitemOpen
  \bibfield  {author} {\bibinfo {author} {\bibfnamefont {Jeffrey~C.Y.}\ \bibnamefont {Teo}}, \bibinfo {author} {\bibfnamefont {Taylor~L.}\ \bibnamefont {Hughes}}, \ and\ \bibinfo {author} {\bibfnamefont {Eduardo}\ \bibnamefont {Fradkin}},\ }\bibfield  {title} {\enquote {\bibinfo {title} {Theory of twist liquids: Gauging an anyonic symmetry},}\ }\href {\doibase 10.1016/j.aop.2015.05.012} {\bibfield  {journal} {\bibinfo  {journal} {Annals of Physics}\ }\textbf {\bibinfo {volume} {360}},\ \bibinfo {pages} {349–445} (\bibinfo {year} {2015})}\BibitemShut {NoStop}%
\bibitem [{\citenamefont {Tarantino}\ \emph {et~al.}(2016)\citenamefont {Tarantino}, \citenamefont {Lindner},\ and\ \citenamefont {Fidkowski}}]{TarantinoSET2016}%
  \BibitemOpen
  \bibfield  {author} {\bibinfo {author} {\bibfnamefont {Nicolas}\ \bibnamefont {Tarantino}}, \bibinfo {author} {\bibfnamefont {Netanel~H}\ \bibnamefont {Lindner}}, \ and\ \bibinfo {author} {\bibfnamefont {Lukasz}\ \bibnamefont {Fidkowski}},\ }\bibfield  {title} {\enquote {\bibinfo {title} {Symmetry fractionalization and twist defects},}\ }\href {http://dx.doi.org/10.1088/1367-2630/18/3/035006} {\bibfield  {journal} {\bibinfo  {journal} {New J. Phys.}\ }\textbf {\bibinfo {volume} {18}},\ \bibinfo {pages} {035006} (\bibinfo {year} {2016})}\BibitemShut {NoStop}%
\bibitem [{\citenamefont {Wen}\ and\ \citenamefont {Zee}(1992)}]{WenZee1992NeutralSF}%
  \BibitemOpen
  \bibfield  {author} {\bibinfo {author} {\bibfnamefont {Xiao-Gang}\ \bibnamefont {Wen}}\ and\ \bibinfo {author} {\bibfnamefont {A.}~\bibnamefont {Zee}},\ }\bibfield  {title} {\enquote {\bibinfo {title} {Neutral superfluid modes and ``magnetic'' monopoles in multilayered quantum hall systems},}\ }\href {\doibase 10.1103/PhysRevLett.69.1811} {\bibfield  {journal} {\bibinfo  {journal} {Phys. Rev. Lett.}\ }\textbf {\bibinfo {volume} {69}},\ \bibinfo {pages} {1811--1814} (\bibinfo {year} {1992})}\BibitemShut {NoStop}%
\bibitem [{\citenamefont {{Senthil}}\ and\ \citenamefont {{Fisher}}(2000)}]{SenthilFisher2000ElecFrac}%
  \BibitemOpen
  \bibfield  {author} {\bibinfo {author} {\bibfnamefont {T.}~\bibnamefont {{Senthil}}}\ and\ \bibinfo {author} {\bibfnamefont {Matthew P.~A.}\ \bibnamefont {{Fisher}}},\ }\bibfield  {title} {\enquote {\bibinfo {title} {{Z$_{2}$ gauge theory of electron fractionalization in strongly correlated systems}},}\ }\href {\doibase 10.1103/PhysRevB.62.7850} {\bibfield  {journal} {\bibinfo  {journal} {\prb}\ }\textbf {\bibinfo {volume} {62}},\ \bibinfo {pages} {7850--7881} (\bibinfo {year} {2000})},\ \Eprint {http://arxiv.org/abs/cond-mat/9910224} {arXiv:cond-mat/9910224 [cond-mat.str-el]} \BibitemShut {NoStop}%
\bibitem [{\citenamefont {{Demler}}\ \emph {et~al.}(2002)\citenamefont {{Demler}}, \citenamefont {{Nayak}}, \citenamefont {{Kee}}, \citenamefont {{Kim}},\ and\ \citenamefont {{Senthil}}}]{Demler2002}%
  \BibitemOpen
  \bibfield  {author} {\bibinfo {author} {\bibfnamefont {Eugene}\ \bibnamefont {{Demler}}}, \bibinfo {author} {\bibfnamefont {Chetan}\ \bibnamefont {{Nayak}}}, \bibinfo {author} {\bibfnamefont {Hae-Young}\ \bibnamefont {{Kee}}}, \bibinfo {author} {\bibfnamefont {Yong~Baek}\ \bibnamefont {{Kim}}}, \ and\ \bibinfo {author} {\bibfnamefont {T.}~\bibnamefont {{Senthil}}},\ }\bibfield  {title} {\enquote {\bibinfo {title} {{Fractionalization patterns in strongly correlated electron systems: Spin-charge separation and beyond}},}\ }\href {\doibase 10.1103/PhysRevB.65.155103} {\bibfield  {journal} {\bibinfo  {journal} {\prb}\ }\textbf {\bibinfo {volume} {65}},\ \bibinfo {eid} {155103} (\bibinfo {year} {2002})},\ \Eprint {http://arxiv.org/abs/cond-mat/0105446} {arXiv:cond-mat/0105446 [cond-mat.str-el]} \BibitemShut {NoStop}%
\bibitem [{\citenamefont {{Paramekanti}}\ and\ \citenamefont {{Vishwanath}}(2004)}]{Paramekanti2004}%
  \BibitemOpen
  \bibfield  {author} {\bibinfo {author} {\bibfnamefont {Arun}\ \bibnamefont {{Paramekanti}}}\ and\ \bibinfo {author} {\bibfnamefont {Ashvin}\ \bibnamefont {{Vishwanath}}},\ }\bibfield  {title} {\enquote {\bibinfo {title} {{Extending Luttinger's theorem to Z$_{2}$ fractionalized phases of matter}},}\ }\href {\doibase 10.1103/PhysRevB.70.245118} {\bibfield  {journal} {\bibinfo  {journal} {\prb}\ }\textbf {\bibinfo {volume} {70}},\ \bibinfo {eid} {245118} (\bibinfo {year} {2004})},\ \Eprint {http://arxiv.org/abs/cond-mat/0406619} {arXiv:cond-mat/0406619 [cond-mat.str-el]} \BibitemShut {NoStop}%
\bibitem [{\citenamefont {{Tang}}\ and\ \citenamefont {{Wen}}(2013)}]{TangWen2013}%
  \BibitemOpen
  \bibfield  {author} {\bibinfo {author} {\bibfnamefont {Evelyn}\ \bibnamefont {{Tang}}}\ and\ \bibinfo {author} {\bibfnamefont {Xiao-Gang}\ \bibnamefont {{Wen}}},\ }\bibfield  {title} {\enquote {\bibinfo {title} {{Superconductivity with intrinsic topological order induced by pure Coulomb interaction and time-reversal symmetry breaking}},}\ }\href {\doibase 10.1103/PhysRevB.88.195117} {\bibfield  {journal} {\bibinfo  {journal} {\prb}\ }\textbf {\bibinfo {volume} {88}},\ \bibinfo {eid} {195117} (\bibinfo {year} {2013})},\ \Eprint {http://arxiv.org/abs/1306.1528} {arXiv:1306.1528 [cond-mat.str-el]} \BibitemShut {NoStop}%
\bibitem [{\citenamefont {Bonderson}\ and\ \citenamefont {Nayak}(2013)}]{BondersonNayak2013QuasiTopo}%
  \BibitemOpen
  \bibfield  {author} {\bibinfo {author} {\bibfnamefont {Parsa}\ \bibnamefont {Bonderson}}\ and\ \bibinfo {author} {\bibfnamefont {Chetan}\ \bibnamefont {Nayak}},\ }\bibfield  {title} {\enquote {\bibinfo {title} {Quasi-topological phases of matter and topological protection},}\ }\href {\doibase 10.1103/PhysRevB.87.195451} {\bibfield  {journal} {\bibinfo  {journal} {Phys. Rev. B}\ }\textbf {\bibinfo {volume} {87}},\ \bibinfo {pages} {195451} (\bibinfo {year} {2013})}\BibitemShut {NoStop}%
\bibitem [{\citenamefont {{Mong}}\ \emph {et~al.}(2014)\citenamefont {{Mong}}, \citenamefont {{Clarke}}, \citenamefont {{Alicea}}, \citenamefont {{Lindner}}, \citenamefont {{Fendley}}, \citenamefont {{Nayak}}, \citenamefont {{Oreg}}, \citenamefont {{Stern}}, \citenamefont {{Berg}}, \citenamefont {{Shtengel}},\ and\ \citenamefont {{Fisher}}}]{Mong2014}%
  \BibitemOpen
  \bibfield  {author} {\bibinfo {author} {\bibfnamefont {Roger S.~K.}\ \bibnamefont {{Mong}}}, \bibinfo {author} {\bibfnamefont {David~J.}\ \bibnamefont {{Clarke}}}, \bibinfo {author} {\bibfnamefont {Jason}\ \bibnamefont {{Alicea}}}, \bibinfo {author} {\bibfnamefont {Netanel~H.}\ \bibnamefont {{Lindner}}}, \bibinfo {author} {\bibfnamefont {Paul}\ \bibnamefont {{Fendley}}}, \bibinfo {author} {\bibfnamefont {Chetan}\ \bibnamefont {{Nayak}}}, \bibinfo {author} {\bibfnamefont {Yuval}\ \bibnamefont {{Oreg}}}, \bibinfo {author} {\bibfnamefont {Ady}\ \bibnamefont {{Stern}}}, \bibinfo {author} {\bibfnamefont {Erez}\ \bibnamefont {{Berg}}}, \bibinfo {author} {\bibfnamefont {Kirill}\ \bibnamefont {{Shtengel}}}, \ and\ \bibinfo {author} {\bibfnamefont {Matthew P.~A.}\ \bibnamefont {{Fisher}}},\ }\bibfield  {title} {\enquote {\bibinfo {title} {{Universal Topological Quantum Computation from a Superconductor-Abelian Quantum Hall Heterostructure}},}\ }\href {\doibase 10.1103/PhysRevX.4.011036} {\bibfield  {journal}
  {\bibinfo  {journal} {Physical Review X}\ }\textbf {\bibinfo {volume} {4}},\ \bibinfo {eid} {011036} (\bibinfo {year} {2014})},\ \Eprint {http://arxiv.org/abs/1307.4403} {arXiv:1307.4403 [cond-mat.str-el]} \BibitemShut {NoStop}%
\bibitem [{\citenamefont {Gazit}\ \emph {et~al.}(2017)\citenamefont {Gazit}, \citenamefont {Randeria},\ and\ \citenamefont {Vishwanath}}]{Gazit2017}%
  \BibitemOpen
  \bibfield  {author} {\bibinfo {author} {\bibfnamefont {Snir}\ \bibnamefont {Gazit}}, \bibinfo {author} {\bibfnamefont {Mohit}\ \bibnamefont {Randeria}}, \ and\ \bibinfo {author} {\bibfnamefont {Ashvin}\ \bibnamefont {Vishwanath}},\ }\bibfield  {title} {\enquote {\bibinfo {title} {Emergent dirac fermions and broken symmetries in confined and deconfined phases of z2 gauge theories},}\ }\href {\doibase 10.1038/nphys4028} {\bibfield  {journal} {\bibinfo  {journal} {Nature Physics}\ }\textbf {\bibinfo {volume} {13}},\ \bibinfo {pages} {484--490} (\bibinfo {year} {2017})}\BibitemShut {NoStop}%
\bibitem [{\citenamefont {Han}\ \emph {et~al.}(2026)\citenamefont {Han}, \citenamefont {Wang}, \citenamefont {Dong}, \citenamefont {Zaletel},\ and\ \citenamefont {Vishwanath}}]{HanWang2026AnyonSF}%
  \BibitemOpen
  \bibfield  {author} {\bibinfo {author} {\bibfnamefont {Zhaoyu}\ \bibnamefont {Han}}, \bibinfo {author} {\bibfnamefont {Taige}\ \bibnamefont {Wang}}, \bibinfo {author} {\bibfnamefont {Zhihuan}\ \bibnamefont {Dong}}, \bibinfo {author} {\bibfnamefont {Michael~P.}\ \bibnamefont {Zaletel}}, \ and\ \bibinfo {author} {\bibfnamefont {Ashvin}\ \bibnamefont {Vishwanath}},\ }\bibfield  {title} {\enquote {\bibinfo {title} {Anyon superfluidity of excitons in quantum hall bilayers},}\ }\href {\doibase 10.1103/xs7g-dmn8} {\bibfield  {journal} {\bibinfo  {journal} {Phys. Rev. X}\ }\textbf {\bibinfo {volume} {16}},\ \bibinfo {pages} {021044} (\bibinfo {year} {2026})}\BibitemShut {NoStop}%
\bibitem [{\citenamefont {{Wang}}\ and\ \citenamefont {{Zhang}}(2025)}]{WangZhang2026AnyonSF}%
  \BibitemOpen
  \bibfield  {author} {\bibinfo {author} {\bibfnamefont {Taige}\ \bibnamefont {{Wang}}}\ and\ \bibinfo {author} {\bibfnamefont {Ya-Hui}\ \bibnamefont {{Zhang}}},\ }\bibfield  {title} {\enquote {\bibinfo {title} {{Anyon superfluid in trilayer quantum Hall systems}},}\ }\href {\doibase 10.48550/arXiv.2508.00058} {\bibfield  {journal} {\bibinfo  {journal} {arXiv e-prints}\ ,\ \bibinfo {eid} {arXiv:2508.00058}} (\bibinfo {year} {2025})},\ \Eprint {http://arxiv.org/abs/2508.00058} {arXiv:2508.00058 [cond-mat.str-el]} \BibitemShut {NoStop}%
\bibitem [{\citenamefont {{Zhang}}\ \emph {et~al.}(2025)\citenamefont {{Zhang}}, \citenamefont {{Zhang}},\ and\ \citenamefont {{Song}}}]{Song20254e}%
  \BibitemOpen
  \bibfield  {author} {\bibinfo {author} {\bibfnamefont {Lu}~\bibnamefont {{Zhang}}}, \bibinfo {author} {\bibfnamefont {Ya-Hui}\ \bibnamefont {{Zhang}}}, \ and\ \bibinfo {author} {\bibfnamefont {Xue-Yang}\ \bibnamefont {{Song}}},\ }\bibfield  {title} {\enquote {\bibinfo {title} {{Charge-4$e$ Anyon Superconductor from Doping $\text{SU}(4)_1$ chiral spin liquid}},}\ }\href {\doibase 10.48550/arXiv.2508.12370} {\bibfield  {journal} {\bibinfo  {journal} {arXiv e-prints}\ ,\ \bibinfo {eid} {arXiv:2508.12370}} (\bibinfo {year} {2025})},\ \Eprint {http://arxiv.org/abs/2508.12370} {arXiv:2508.12370 [cond-mat.str-el]} \BibitemShut {NoStop}%
\bibitem [{\citenamefont {{Darius Shi}}\ and\ \citenamefont {{Senthil}}(2025)}]{ShiSenthil2025NonAbelianSC}%
  \BibitemOpen
  \bibfield  {author} {\bibinfo {author} {\bibfnamefont {Zhengyan}\ \bibnamefont {{Darius Shi}}}\ and\ \bibinfo {author} {\bibfnamefont {T.}~\bibnamefont {{Senthil}}},\ }\bibfield  {title} {\enquote {\bibinfo {title} {{Non-Abelian topological superconductivity from melting Abelian fractional Chern insulators}},}\ }\href {\doibase 10.48550/arXiv.2512.17996} {\bibfield  {journal} {\bibinfo  {journal} {arXiv e-prints}\ ,\ \bibinfo {eid} {arXiv:2512.17996}} (\bibinfo {year} {2025})},\ \Eprint {http://arxiv.org/abs/2512.17996} {arXiv:2512.17996 [cond-mat.str-el]} \BibitemShut {NoStop}%
\bibitem [{\citenamefont {Shi}\ \emph {et~al.}(2025)\citenamefont {Shi}, \citenamefont {Zhang},\ and\ \citenamefont {Todadri}}]{ShiZhangSenthil2025}%
  \BibitemOpen
  \bibfield  {author} {\bibinfo {author} {\bibfnamefont {Zhengyan~Darius}\ \bibnamefont {Shi}}, \bibinfo {author} {\bibfnamefont {Carolyn}\ \bibnamefont {Zhang}}, \ and\ \bibinfo {author} {\bibfnamefont {Senthil}\ \bibnamefont {Todadri}},\ }\bibfield  {title} {\enquote {\bibinfo {title} {Doping lattice non-abelian quantum hall states},}\ }\href {\doibase 10.21468/SciPostPhys.19.6.150} {\bibfield  {journal} {\bibinfo  {journal} {SciPost Phys.}\ }\textbf {\bibinfo {volume} {19}},\ \bibinfo {pages} {150} (\bibinfo {year} {2025})}\BibitemShut {NoStop}%
\bibitem [{\citenamefont {{Gao}}\ \emph {et~al.}(2025)\citenamefont {{Gao}}, \citenamefont {{Wang}}, \citenamefont {{Yang}},\ and\ \citenamefont {{Wu}}}]{GaoWu2025}%
  \BibitemOpen
  \bibfield  {author} {\bibinfo {author} {\bibfnamefont {Zhi-Qiang}\ \bibnamefont {{Gao}}}, \bibinfo {author} {\bibfnamefont {Yan-Qi}\ \bibnamefont {{Wang}}}, \bibinfo {author} {\bibfnamefont {Hui}\ \bibnamefont {{Yang}}}, \ and\ \bibinfo {author} {\bibfnamefont {Congjun}\ \bibnamefont {{Wu}}},\ }\bibfield  {title} {\enquote {\bibinfo {title} {{Topological Charge-2ne Superconductors}},}\ }\href {\doibase 10.48550/arXiv.2512.21325} {\bibfield  {journal} {\bibinfo  {journal} {arXiv e-prints}\ ,\ \bibinfo {eid} {arXiv:2512.21325}} (\bibinfo {year} {2025})},\ \Eprint {http://arxiv.org/abs/2512.21325} {arXiv:2512.21325 [cond-mat.str-el]} \BibitemShut {NoStop}%
\bibitem [{\citenamefont {{Wang}}\ \emph {et~al.}(2026)\citenamefont {{Wang}}, \citenamefont {{Zhou}}, \citenamefont {{Yang}},\ and\ \citenamefont {{Gao}}}]{Gao2026}%
  \BibitemOpen
  \bibfield  {author} {\bibinfo {author} {\bibfnamefont {Yan-Qi}\ \bibnamefont {{Wang}}}, \bibinfo {author} {\bibfnamefont {Boran}\ \bibnamefont {{Zhou}}}, \bibinfo {author} {\bibfnamefont {Hui}\ \bibnamefont {{Yang}}}, \ and\ \bibinfo {author} {\bibfnamefont {Zhi-Qiang}\ \bibnamefont {{Gao}}},\ }\bibfield  {title} {\enquote {\bibinfo {title} {{Charge-6e superconductivity from doping SU(3) spin liquids}},}\ }\href {\doibase 10.48550/arXiv.2607.25909} {\bibfield  {journal} {\bibinfo  {journal} {arXiv e-prints}\ ,\ \bibinfo {eid} {arXiv:2607.25909}} (\bibinfo {year} {2026})},\ \Eprint {http://arxiv.org/abs/2607.25909} {arXiv:2607.25909 [cond-mat.str-el]} \BibitemShut {NoStop}%
\bibitem [{\citenamefont {{Darius Shi}}\ \emph {et~al.}(2026)\citenamefont {{Darius Shi}}, \citenamefont {{Han}}, \citenamefont {{Raghu}},\ and\ \citenamefont {{Vishwanath}}}]{ShiHan4e2026}%
  \BibitemOpen
  \bibfield  {author} {\bibinfo {author} {\bibfnamefont {Zhengyan}\ \bibnamefont {{Darius Shi}}}, \bibinfo {author} {\bibfnamefont {Zhaoyu}\ \bibnamefont {{Han}}}, \bibinfo {author} {\bibfnamefont {Srinivas}\ \bibnamefont {{Raghu}}}, \ and\ \bibinfo {author} {\bibfnamefont {Ashvin}\ \bibnamefont {{Vishwanath}}},\ }\bibfield  {title} {\enquote {\bibinfo {title} {{Charge-$4e$ superconductor with parafermionic vortices: A path to universal topological quantum computation}},}\ }\href {\doibase 10.48550/arXiv.2602.06963} {\bibfield  {journal} {\bibinfo  {journal} {arXiv e-prints}\ ,\ \bibinfo {eid} {arXiv:2602.06963}} (\bibinfo {year} {2026})},\ \Eprint {http://arxiv.org/abs/2602.06963} {arXiv:2602.06963 [cond-mat.str-el]} \BibitemShut {NoStop}%
\bibitem [{\citenamefont {Shi}\ and\ \citenamefont {Senthil}(2025)}]{ShiSenthil2025}%
  \BibitemOpen
  \bibfield  {author} {\bibinfo {author} {\bibfnamefont {Zhengyan~Darius}\ \bibnamefont {Shi}}\ and\ \bibinfo {author} {\bibfnamefont {T.}~\bibnamefont {Senthil}},\ }\bibfield  {title} {\enquote {\bibinfo {title} {Doping a fractional quantum anomalous hall insulator},}\ }\href {\doibase 10.1103/kcm5-hx56} {\bibfield  {journal} {\bibinfo  {journal} {Phys. Rev. X}\ }\textbf {\bibinfo {volume} {15}},\ \bibinfo {pages} {031069} (\bibinfo {year} {2025})}\BibitemShut {NoStop}%
\bibitem [{\citenamefont {{Cheng}}\ and\ \citenamefont {{Jian}}(2022)}]{GaugingU1}%
  \BibitemOpen
  \bibfield  {author} {\bibinfo {author} {\bibfnamefont {Meng}\ \bibnamefont {{Cheng}}}\ and\ \bibinfo {author} {\bibfnamefont {Chao-Ming}\ \bibnamefont {{Jian}}},\ }\bibfield  {title} {\enquote {\bibinfo {title} {{Gauging U(1) symmetry in (2+1)d topological phases}},}\ }\href {\doibase 10.21468/SciPostPhys.12.6.202} {\bibfield  {journal} {\bibinfo  {journal} {SciPost Physics}\ }\textbf {\bibinfo {volume} {12}},\ \bibinfo {eid} {202} (\bibinfo {year} {2022})},\ \Eprint {http://arxiv.org/abs/2201.07239} {arXiv:2201.07239 [cond-mat.str-el]} \BibitemShut {NoStop}%
\bibitem [{\citenamefont {{Else}}(2021)}]{Else2021TopoGoldstone}%
  \BibitemOpen
  \bibfield  {author} {\bibinfo {author} {\bibfnamefont {Dominic~V.}\ \bibnamefont {{Else}}},\ }\bibfield  {title} {\enquote {\bibinfo {title} {{Topological Goldstone phases of matter}},}\ }\href {\doibase 10.1103/PhysRevB.104.115129} {\bibfield  {journal} {\bibinfo  {journal} {\prb}\ }\textbf {\bibinfo {volume} {104}},\ \bibinfo {eid} {115129} (\bibinfo {year} {2021})},\ \Eprint {http://arxiv.org/abs/2102.08953} {arXiv:2102.08953 [cond-mat.str-el]} \BibitemShut {NoStop}%
\bibitem [{\citenamefont {{Cheng}}\ \emph {et~al.}(2023)\citenamefont {{Cheng}}, \citenamefont {{Hsin}},\ and\ \citenamefont {{Jian}}}]{GaugingLieGroup}%
  \BibitemOpen
  \bibfield  {author} {\bibinfo {author} {\bibfnamefont {Meng}\ \bibnamefont {{Cheng}}}, \bibinfo {author} {\bibfnamefont {Po-Shen}\ \bibnamefont {{Hsin}}}, \ and\ \bibinfo {author} {\bibfnamefont {Chao-Ming}\ \bibnamefont {{Jian}}},\ }\bibfield  {title} {\enquote {\bibinfo {title} {{Gauging Lie group symmetry in (2+1)d topological phases}},}\ }\href {\doibase 10.21468/SciPostPhys.14.5.100} {\bibfield  {journal} {\bibinfo  {journal} {SciPost Physics}\ }\textbf {\bibinfo {volume} {14}},\ \bibinfo {eid} {100} (\bibinfo {year} {2023})},\ \Eprint {http://arxiv.org/abs/2205.15347} {arXiv:2205.15347 [cond-mat.str-el]} \BibitemShut {NoStop}%
\bibitem [{Note1()}]{Note1}%
  \BibitemOpen
  \bibinfo {note} {The sign convention for the Hall conductance $\sigma _{xy}$ is fixed by the response Lagrangian $ {\protect \cal L} _{\protect \rm response} = -\protect \frac {\sigma _{xy}}{4\pi }BdB$, where $B$ is a background $\protect \mathrm {U}(1)$ gauge field.}\BibitemShut {Stop}%
\bibitem [{\citenamefont {{Bais}}\ and\ \citenamefont {{Slingerland}}(2009)}]{BaisSlingerland2009}%
  \BibitemOpen
  \bibfield  {author} {\bibinfo {author} {\bibfnamefont {F.~A.}\ \bibnamefont {{Bais}}}\ and\ \bibinfo {author} {\bibfnamefont {J.~K.}\ \bibnamefont {{Slingerland}}},\ }\bibfield  {title} {\enquote {\bibinfo {title} {{Condensate-induced transitions between topologically ordered phases}},}\ }\href {\doibase 10.1103/PhysRevB.79.045316} {\bibfield  {journal} {\bibinfo  {journal} {\prb}\ }\textbf {\bibinfo {volume} {79}},\ \bibinfo {eid} {045316} (\bibinfo {year} {2009})},\ \Eprint {http://arxiv.org/abs/0808.0627} {arXiv:0808.0627 [cond-mat.mes-hall]} \BibitemShut {NoStop}%
\bibitem [{Note2()}]{Note2}%
  \BibitemOpen
  \bibinfo {note} {In our notation, the superscript ``${}^{(k)}$'' emphasizes that the symmetry is a $k$-form symmetry. The $\protect \mathrm {U}(1)$ symmetries in this work are all 0-form symmetries, so we suppress the superscript.}\BibitemShut {Stop}%
\bibitem [{\citenamefont {Gaiotto}\ \emph {et~al.}(2015)\citenamefont {Gaiotto}, \citenamefont {Kapustin}, \citenamefont {Seiberg},\ and\ \citenamefont {Willett}}]{Gaiotto2015}%
  \BibitemOpen
  \bibfield  {author} {\bibinfo {author} {\bibfnamefont {Davide}\ \bibnamefont {Gaiotto}}, \bibinfo {author} {\bibfnamefont {Anton}\ \bibnamefont {Kapustin}}, \bibinfo {author} {\bibfnamefont {Nathan}\ \bibnamefont {Seiberg}}, \ and\ \bibinfo {author} {\bibfnamefont {Brian}\ \bibnamefont {Willett}},\ }\bibfield  {title} {\enquote {\bibinfo {title} {Generalized global symmetries},}\ }\href {\doibase 10.1007/jhep02(2015)172} {\bibfield  {journal} {\bibinfo  {journal} {J. High Energ. Phys.}\ }\textbf {\bibinfo {volume} {2015}} (\bibinfo {year} {2015}),\ 10.1007/jhep02(2015)172}\BibitemShut {NoStop}%
\bibitem [{\citenamefont {Moore}\ and\ \citenamefont {Seiberg}(1989)}]{Moore:1989yh}%
  \BibitemOpen
  \bibfield  {author} {\bibinfo {author} {\bibfnamefont {Gregory~W.}\ \bibnamefont {Moore}}\ and\ \bibinfo {author} {\bibfnamefont {Nathan}\ \bibnamefont {Seiberg}},\ }\bibfield  {title} {\enquote {\bibinfo {title} {{Taming the Conformal Zoo}},}\ }\href {\doibase 10.1016/0370-2693(89)90897-6} {\bibfield  {journal} {\bibinfo  {journal} {Phys. Lett. B}\ }\textbf {\bibinfo {volume} {220}},\ \bibinfo {pages} {422--430} (\bibinfo {year} {1989})}\BibitemShut {NoStop}%
\bibitem [{\citenamefont {Bombin}(2010)}]{Bombin2010_dislocation}%
  \BibitemOpen
  \bibfield  {author} {\bibinfo {author} {\bibfnamefont {H.}~\bibnamefont {Bombin}},\ }\bibfield  {title} {\enquote {\bibinfo {title} {Topological order with a twist: Ising anyons from an abelian model},}\ }\href {\doibase 10.1103/PhysRevLett.105.030403} {\bibfield  {journal} {\bibinfo  {journal} {Phys. Rev. Lett.}\ }\textbf {\bibinfo {volume} {105}},\ \bibinfo {pages} {030403} (\bibinfo {year} {2010})}\BibitemShut {NoStop}%
\bibitem [{\citenamefont {{Lindner}}\ \emph {et~al.}(2012)\citenamefont {{Lindner}}, \citenamefont {{Berg}}, \citenamefont {{Refael}},\ and\ \citenamefont {{Stern}}}]{Lindner2012}%
  \BibitemOpen
  \bibfield  {author} {\bibinfo {author} {\bibfnamefont {Netanel~H.}\ \bibnamefont {{Lindner}}}, \bibinfo {author} {\bibfnamefont {Erez}\ \bibnamefont {{Berg}}}, \bibinfo {author} {\bibfnamefont {Gil}\ \bibnamefont {{Refael}}}, \ and\ \bibinfo {author} {\bibfnamefont {Ady}\ \bibnamefont {{Stern}}},\ }\bibfield  {title} {\enquote {\bibinfo {title} {{Fractionalizing Majorana Fermions: Non-Abelian Statistics on the Edges of Abelian Quantum Hall States}},}\ }\href {\doibase 10.1103/PhysRevX.2.041002} {\bibfield  {journal} {\bibinfo  {journal} {Physical Review X}\ }\textbf {\bibinfo {volume} {2}},\ \bibinfo {eid} {041002} (\bibinfo {year} {2012})},\ \Eprint {http://arxiv.org/abs/1204.5733} {arXiv:1204.5733 [cond-mat.mes-hall]} \BibitemShut {NoStop}%
\bibitem [{\citenamefont {{Cheng}}(2012)}]{Meng2012}%
  \BibitemOpen
  \bibfield  {author} {\bibinfo {author} {\bibfnamefont {Meng}\ \bibnamefont {{Cheng}}},\ }\bibfield  {title} {\enquote {\bibinfo {title} {{Superconducting proximity effect on the edge of fractional topological insulators}},}\ }\href {\doibase 10.1103/PhysRevB.86.195126} {\bibfield  {journal} {\bibinfo  {journal} {\prb}\ }\textbf {\bibinfo {volume} {86}},\ \bibinfo {eid} {195126} (\bibinfo {year} {2012})},\ \Eprint {http://arxiv.org/abs/1204.6084} {arXiv:1204.6084 [cond-mat.str-el]} \BibitemShut {NoStop}%
\bibitem [{\citenamefont {{Clarke}}\ \emph {et~al.}(2013)\citenamefont {{Clarke}}, \citenamefont {{Alicea}},\ and\ \citenamefont {{Shtengel}}}]{Clarke2013}%
  \BibitemOpen
  \bibfield  {author} {\bibinfo {author} {\bibfnamefont {David~J.}\ \bibnamefont {{Clarke}}}, \bibinfo {author} {\bibfnamefont {Jason}\ \bibnamefont {{Alicea}}}, \ and\ \bibinfo {author} {\bibfnamefont {Kirill}\ \bibnamefont {{Shtengel}}},\ }\bibfield  {title} {\enquote {\bibinfo {title} {{Exotic non-Abelian anyons from conventional fractional quantum Hall states}},}\ }\href {\doibase 10.1038/ncomms2340} {\bibfield  {journal} {\bibinfo  {journal} {Nature Communications}\ }\textbf {\bibinfo {volume} {4}},\ \bibinfo {eid} {1348} (\bibinfo {year} {2013})},\ \Eprint {http://arxiv.org/abs/1204.5479} {arXiv:1204.5479 [cond-mat.str-el]} \BibitemShut {NoStop}%
\bibitem [{\citenamefont {{You}}\ and\ \citenamefont {{Wen}}(2012)}]{YouWen2012}%
  \BibitemOpen
  \bibfield  {author} {\bibinfo {author} {\bibfnamefont {Yi-Zhuang}\ \bibnamefont {{You}}}\ and\ \bibinfo {author} {\bibfnamefont {Xiao-Gang}\ \bibnamefont {{Wen}}},\ }\bibfield  {title} {\enquote {\bibinfo {title} {{Projective non-Abelian statistics of dislocation defects in a Z$_{N}$ rotor model}},}\ }\href {\doibase 10.1103/PhysRevB.86.161107} {\bibfield  {journal} {\bibinfo  {journal} {\prb}\ }\textbf {\bibinfo {volume} {86}},\ \bibinfo {eid} {161107} (\bibinfo {year} {2012})},\ \Eprint {http://arxiv.org/abs/1204.0113} {arXiv:1204.0113 [cond-mat.str-el]} \BibitemShut {NoStop}%
\bibitem [{\citenamefont {{Barkeshli}}\ and\ \citenamefont {{Qi}}(2012)}]{BarkeshliQi2012}%
  \BibitemOpen
  \bibfield  {author} {\bibinfo {author} {\bibfnamefont {Maissam}\ \bibnamefont {{Barkeshli}}}\ and\ \bibinfo {author} {\bibfnamefont {Xiao-Liang}\ \bibnamefont {{Qi}}},\ }\bibfield  {title} {\enquote {\bibinfo {title} {{Topological Nematic States and Non-Abelian Lattice Dislocations}},}\ }\href {\doibase 10.1103/PhysRevX.2.031013} {\bibfield  {journal} {\bibinfo  {journal} {Physical Review X}\ }\textbf {\bibinfo {volume} {2}},\ \bibinfo {eid} {031013} (\bibinfo {year} {2012})},\ \Eprint {http://arxiv.org/abs/1112.3311} {arXiv:1112.3311 [cond-mat.str-el]} \BibitemShut {NoStop}%
\bibitem [{\citenamefont {{Barkeshli}}\ \emph {et~al.}(2013)\citenamefont {{Barkeshli}}, \citenamefont {{Jian}},\ and\ \citenamefont {{Qi}}}]{Barkeshli2012_genon}%
  \BibitemOpen
  \bibfield  {author} {\bibinfo {author} {\bibfnamefont {Maissam}\ \bibnamefont {{Barkeshli}}}, \bibinfo {author} {\bibfnamefont {Chao-Ming}\ \bibnamefont {{Jian}}}, \ and\ \bibinfo {author} {\bibfnamefont {Xiao-Liang}\ \bibnamefont {{Qi}}},\ }\bibfield  {title} {\enquote {\bibinfo {title} {{Twist defects and projective non-Abelian braiding statistics}},}\ }\href {\doibase 10.1103/PhysRevB.87.045130} {\bibfield  {journal} {\bibinfo  {journal} {\prb}\ }\textbf {\bibinfo {volume} {87}},\ \bibinfo {eid} {045130} (\bibinfo {year} {2013})},\ \Eprint {http://arxiv.org/abs/1208.4834} {arXiv:1208.4834 [cond-mat.str-el]} \BibitemShut {NoStop}%
\bibitem [{\citenamefont {You}\ \emph {et~al.}(2013)\citenamefont {You}, \citenamefont {Jian},\ and\ \citenamefont {Wen}}]{YouJianWen2013}%
  \BibitemOpen
  \bibfield  {author} {\bibinfo {author} {\bibfnamefont {Yi-Zhuang}\ \bibnamefont {You}}, \bibinfo {author} {\bibfnamefont {Chao-Ming}\ \bibnamefont {Jian}}, \ and\ \bibinfo {author} {\bibfnamefont {Xiao-Gang}\ \bibnamefont {Wen}},\ }\bibfield  {title} {\enquote {\bibinfo {title} {Synthetic non-abelian statistics by abelian anyon condensation},}\ }\href {\doibase 10.1103/PhysRevB.87.045106} {\bibfield  {journal} {\bibinfo  {journal} {Phys. Rev. B}\ }\textbf {\bibinfo {volume} {87}},\ \bibinfo {pages} {045106} (\bibinfo {year} {2013})}\BibitemShut {NoStop}%
\bibitem [{\citenamefont {Yoder}\ and\ \citenamefont {Kim}(2017)}]{Yoder2017_surfacecodetwist}%
  \BibitemOpen
  \bibfield  {author} {\bibinfo {author} {\bibfnamefont {Theodore~J.}\ \bibnamefont {Yoder}}\ and\ \bibinfo {author} {\bibfnamefont {Isaac~H.}\ \bibnamefont {Kim}},\ }\bibfield  {title} {\enquote {\bibinfo {title} {The surface code with a twist},}\ }\href {\doibase 10.22331/q-2017-04-25-2} {\bibfield  {journal} {\bibinfo  {journal} {{Quantum}}\ }\textbf {\bibinfo {volume} {1}},\ \bibinfo {pages} {2} (\bibinfo {year} {2017})}\BibitemShut {NoStop}%
\bibitem [{\citenamefont {{Kesselring}}\ \emph {et~al.}(2018)\citenamefont {{Kesselring}}, \citenamefont {{Pastawski}}, \citenamefont {{Eisert}},\ and\ \citenamefont {{Brown}}}]{kesselring2018}%
  \BibitemOpen
  \bibfield  {author} {\bibinfo {author} {\bibfnamefont {Markus~S.}\ \bibnamefont {{Kesselring}}}, \bibinfo {author} {\bibfnamefont {Fernando}\ \bibnamefont {{Pastawski}}}, \bibinfo {author} {\bibfnamefont {Jens}\ \bibnamefont {{Eisert}}}, \ and\ \bibinfo {author} {\bibfnamefont {Benjamin~J.}\ \bibnamefont {{Brown}}},\ }\bibfield  {title} {\enquote {\bibinfo {title} {{The boundaries and twist defects of the color code and their applications to topological quantum computation}},}\ }\href {\doibase 10.22331/q-2018-10-19-101} {\bibfield  {journal} {\bibinfo  {journal} {Quantum}\ }\textbf {\bibinfo {volume} {2}},\ \bibinfo {pages} {101} (\bibinfo {year} {2018})},\ \Eprint {http://arxiv.org/abs/1806.02820} {arXiv:1806.02820 [quant-ph]} \BibitemShut {NoStop}%
\bibitem [{\citenamefont {Teo}(2016)}]{Teo_2016}%
  \BibitemOpen
  \bibfield  {author} {\bibinfo {author} {\bibfnamefont {Jeffrey C~Y}\ \bibnamefont {Teo}},\ }\bibfield  {title} {\enquote {\bibinfo {title} {Globally symmetric topological phase: from anyonic symmetry to twist defect},}\ }\href {\doibase 10.1088/0953-8984/28/14/143001} {\bibfield  {journal} {\bibinfo  {journal} {Journal of Physics: Condensed Matter}\ }\textbf {\bibinfo {volume} {28}},\ \bibinfo {pages} {143001} (\bibinfo {year} {2016})}\BibitemShut {NoStop}%
\bibitem [{\citenamefont {Senthil}\ and\ \citenamefont {Levin}(2013)}]{BIQH2013}%
  \BibitemOpen
  \bibfield  {author} {\bibinfo {author} {\bibfnamefont {T.}~\bibnamefont {Senthil}}\ and\ \bibinfo {author} {\bibfnamefont {Michael}\ \bibnamefont {Levin}},\ }\bibfield  {title} {\enquote {\bibinfo {title} {Integer quantum hall effect for bosons},}\ }\href {\doibase 10.1103/PhysRevLett.110.046801} {\bibfield  {journal} {\bibinfo  {journal} {Phys. Rev. Lett.}\ }\textbf {\bibinfo {volume} {110}},\ \bibinfo {pages} {046801} (\bibinfo {year} {2013})}\BibitemShut {NoStop}%
\bibitem [{\citenamefont {{Read}}\ and\ \citenamefont {{Rezayi}}(1999)}]{ReadRezayi1999}%
  \BibitemOpen
  \bibfield  {author} {\bibinfo {author} {\bibfnamefont {N.}~\bibnamefont {{Read}}}\ and\ \bibinfo {author} {\bibfnamefont {E.}~\bibnamefont {{Rezayi}}},\ }\bibfield  {title} {\enquote {\bibinfo {title} {{Beyond paired quantum Hall states: Parafermions and incompressible states in the first excited Landau level}},}\ }\href {\doibase 10.1103/PhysRevB.59.8084} {\bibfield  {journal} {\bibinfo  {journal} {\prb}\ }\textbf {\bibinfo {volume} {59}},\ \bibinfo {pages} {8084--8092} (\bibinfo {year} {1999})},\ \Eprint {http://arxiv.org/abs/cond-mat/9809384} {arXiv:cond-mat/9809384 [cond-mat.mes-hall]} \BibitemShut {NoStop}%
\bibitem [{\citenamefont {{Cappelli}}\ \emph {et~al.}(2001)\citenamefont {{Cappelli}}, \citenamefont {{Georgiev}},\ and\ \citenamefont {{Todorov}}}]{Cappelli2001_para}%
  \BibitemOpen
  \bibfield  {author} {\bibinfo {author} {\bibfnamefont {Andrea}\ \bibnamefont {{Cappelli}}}, \bibinfo {author} {\bibfnamefont {Lachezar~S.}\ \bibnamefont {{Georgiev}}}, \ and\ \bibinfo {author} {\bibfnamefont {Ivan~T.}\ \bibnamefont {{Todorov}}},\ }\bibfield  {title} {\enquote {\bibinfo {title} {{Parafermion Hall states from coset projections of abelian conformal theories}},}\ }\href {\doibase 10.1016/S0550-3213(00)00774-4} {\bibfield  {journal} {\bibinfo  {journal} {Nuclear Physics B}\ }\textbf {\bibinfo {volume} {599}},\ \bibinfo {pages} {499--530} (\bibinfo {year} {2001})},\ \Eprint {http://arxiv.org/abs/hep-th/0009229} {arXiv:hep-th/0009229 [hep-th]} \BibitemShut {NoStop}%
\bibitem [{\citenamefont {{Barkeshli}}\ and\ \citenamefont {{Wen}}(2010)}]{BarkeshliWen2010_projective}%
  \BibitemOpen
  \bibfield  {author} {\bibinfo {author} {\bibfnamefont {Maissam}\ \bibnamefont {{Barkeshli}}}\ and\ \bibinfo {author} {\bibfnamefont {Xiao-Gang}\ \bibnamefont {{Wen}}},\ }\bibfield  {title} {\enquote {\bibinfo {title} {{Effective field theory and projective construction for Z$_{k}$ parafermion fractional quantum Hall states}},}\ }\href {\doibase 10.1103/PhysRevB.81.155302} {\bibfield  {journal} {\bibinfo  {journal} {\prb}\ }\textbf {\bibinfo {volume} {81}},\ \bibinfo {eid} {155302} (\bibinfo {year} {2010})},\ \Eprint {http://arxiv.org/abs/0910.2483} {arXiv:0910.2483 [cond-mat.str-el]} \BibitemShut {NoStop}%
\bibitem [{\citenamefont {{Hsin}}\ and\ \citenamefont {{Seiberg}}(2016)}]{Hsin2016LevelRankDuality}%
  \BibitemOpen
  \bibfield  {author} {\bibinfo {author} {\bibfnamefont {Po-Shen}\ \bibnamefont {{Hsin}}}\ and\ \bibinfo {author} {\bibfnamefont {Nathan}\ \bibnamefont {{Seiberg}}},\ }\bibfield  {title} {\enquote {\bibinfo {title} {{Level/rank duality and Chern-Simons-matter theories}},}\ }\href {\doibase 10.1007/JHEP09(2016)095} {\bibfield  {journal} {\bibinfo  {journal} {Journal of High Energy Physics}\ }\textbf {\bibinfo {volume} {2016}},\ \bibinfo {eid} {95} (\bibinfo {year} {2016})},\ \Eprint {http://arxiv.org/abs/1607.07457} {arXiv:1607.07457 [hep-th]} \BibitemShut {NoStop}%
\bibitem [{\citenamefont {{Ardonne}}\ and\ \citenamefont {{Schoutens}}(1999)}]{Ardonne1999}%
  \BibitemOpen
  \bibfield  {author} {\bibinfo {author} {\bibfnamefont {Eddy}\ \bibnamefont {{Ardonne}}}\ and\ \bibinfo {author} {\bibfnamefont {Kareljan}\ \bibnamefont {{Schoutens}}},\ }\bibfield  {title} {\enquote {\bibinfo {title} {{New Class of Non-Abelian Spin-Singlet Quantum Hall States}},}\ }\href {\doibase 10.1103/PhysRevLett.82.5096} {\bibfield  {journal} {\bibinfo  {journal} {\prl}\ }\textbf {\bibinfo {volume} {82}},\ \bibinfo {pages} {5096--5099} (\bibinfo {year} {1999})},\ \Eprint {http://arxiv.org/abs/cond-mat/9811352} {arXiv:cond-mat/9811352 [cond-mat.mes-hall]} \BibitemShut {NoStop}%
\bibitem [{\citenamefont {{Ardonne}}\ \emph {et~al.}(2001)\citenamefont {{Ardonne}}, \citenamefont {{Read}}, \citenamefont {{Rezayi}},\ and\ \citenamefont {{Schoutens}}}]{Ardonne2001}%
  \BibitemOpen
  \bibfield  {author} {\bibinfo {author} {\bibfnamefont {E.}~\bibnamefont {{Ardonne}}}, \bibinfo {author} {\bibfnamefont {N.}~\bibnamefont {{Read}}}, \bibinfo {author} {\bibfnamefont {E.}~\bibnamefont {{Rezayi}}}, \ and\ \bibinfo {author} {\bibfnamefont {K.}~\bibnamefont {{Schoutens}}},\ }\bibfield  {title} {\enquote {\bibinfo {title} {{Non-abelian spin-singlet quantum Hall states: wave functions and quasihole state counting}},}\ }\href {\doibase 10.1016/S0550-3213(01)00224-3} {\bibfield  {journal} {\bibinfo  {journal} {Nuclear Physics B}\ }\textbf {\bibinfo {volume} {607}},\ \bibinfo {pages} {549--576} (\bibinfo {year} {2001})},\ \Eprint {http://arxiv.org/abs/cond-mat/0104250} {arXiv:cond-mat/0104250 [cond-mat.mes-hall]} \BibitemShut {NoStop}%
\bibitem [{\citenamefont {Dasgupta}\ and\ \citenamefont {Halperin}(1981)}]{Dasgupta1981_latticePV}%
  \BibitemOpen
  \bibfield  {author} {\bibinfo {author} {\bibfnamefont {C.}~\bibnamefont {Dasgupta}}\ and\ \bibinfo {author} {\bibfnamefont {B.~I.}\ \bibnamefont {Halperin}},\ }\bibfield  {title} {\enquote {\bibinfo {title} {Phase transition in a lattice model of superconductivity},}\ }\href {\doibase 10.1103/PhysRevLett.47.1556} {\bibfield  {journal} {\bibinfo  {journal} {Phys. Rev. Lett.}\ }\textbf {\bibinfo {volume} {47}},\ \bibinfo {pages} {1556--1560} (\bibinfo {year} {1981})}\BibitemShut {NoStop}%
\bibitem [{\citenamefont {{Peskin}}(1978)}]{Peskin1978_latticePV}%
  \BibitemOpen
  \bibfield  {author} {\bibinfo {author} {\bibfnamefont {Michael~E.}\ \bibnamefont {{Peskin}}},\ }\bibfield  {title} {\enquote {\bibinfo {title} {{Mandelstam-'t Hooft duality in abelian lattice models}},}\ }\href {\doibase 10.1016/0003-4916(78)90252-X} {\bibfield  {journal} {\bibinfo  {journal} {Annals of Physics}\ }\textbf {\bibinfo {volume} {113}},\ \bibinfo {pages} {122--152} (\bibinfo {year} {1978})}\BibitemShut {NoStop}%
\bibitem [{\citenamefont {{Cheng}}\ and\ \citenamefont {{Seiberg}}(2026)}]{ChengSeiberg2026}%
  \BibitemOpen
  \bibfield  {author} {\bibinfo {author} {\bibfnamefont {Meng}\ \bibnamefont {{Cheng}}}\ and\ \bibinfo {author} {\bibfnamefont {Nathan}\ \bibnamefont {{Seiberg}}},\ }\bibfield  {title} {\enquote {\bibinfo {title} {{Proliferation transitions from a topological phase in 2+1 dimensions}},}\ }\href {\doibase 10.1103/bfzc-7lm2} {\bibfield  {journal} {\bibinfo  {journal} {\prb}\ }\textbf {\bibinfo {volume} {114}},\ \bibinfo {eid} {055103} (\bibinfo {year} {2026})},\ \Eprint {http://arxiv.org/abs/2603.00245} {arXiv:2603.00245 [cond-mat.str-el]} \BibitemShut {NoStop}%
\bibitem [{\citenamefont {{Read}}\ and\ \citenamefont {{Green}}(2000)}]{ReadGreen2000}%
  \BibitemOpen
  \bibfield  {author} {\bibinfo {author} {\bibfnamefont {N.}~\bibnamefont {{Read}}}\ and\ \bibinfo {author} {\bibfnamefont {Dmitry}\ \bibnamefont {{Green}}},\ }\bibfield  {title} {\enquote {\bibinfo {title} {{Paired states of fermions in two dimensions with breaking of parity and time-reversal symmetries and the fractional quantum Hall effect}},}\ }\href {\doibase 10.1103/PhysRevB.61.10267} {\bibfield  {journal} {\bibinfo  {journal} {\prb}\ }\textbf {\bibinfo {volume} {61}},\ \bibinfo {pages} {10267--10297} (\bibinfo {year} {2000})},\ \Eprint {http://arxiv.org/abs/cond-mat/9906453} {arXiv:cond-mat/9906453 [cond-mat.mes-hall]} \BibitemShut {NoStop}%
\bibitem [{\citenamefont {{Ivanov}}(2001)}]{Ivanov2001}%
  \BibitemOpen
  \bibfield  {author} {\bibinfo {author} {\bibfnamefont {D.~A.}\ \bibnamefont {{Ivanov}}},\ }\bibfield  {title} {\enquote {\bibinfo {title} {{Non-Abelian Statistics of Half-Quantum Vortices in p-Wave Superconductors}},}\ }\href {\doibase 10.1103/PhysRevLett.86.268} {\bibfield  {journal} {\bibinfo  {journal} {\prl}\ }\textbf {\bibinfo {volume} {86}},\ \bibinfo {pages} {268--271} (\bibinfo {year} {2001})},\ \Eprint {http://arxiv.org/abs/cond-mat/0005069} {arXiv:cond-mat/0005069 [cond-mat.supr-con]} \BibitemShut {NoStop}%
\bibitem [{\citenamefont {Sondhi}\ \emph {et~al.}(1993)\citenamefont {Sondhi}, \citenamefont {Karlhede}, \citenamefont {Kivelson},\ and\ \citenamefont {Rezayi}}]{Sondhi1993}%
  \BibitemOpen
  \bibfield  {author} {\bibinfo {author} {\bibfnamefont {S.~L.}\ \bibnamefont {Sondhi}}, \bibinfo {author} {\bibfnamefont {A.}~\bibnamefont {Karlhede}}, \bibinfo {author} {\bibfnamefont {S.~A.}\ \bibnamefont {Kivelson}}, \ and\ \bibinfo {author} {\bibfnamefont {E.~H.}\ \bibnamefont {Rezayi}},\ }\bibfield  {title} {\enquote {\bibinfo {title} {Skyrmions and the crossover from the integer to fractional quantum hall effect at small zeeman energies},}\ }\href {\doibase 10.1103/PhysRevB.47.16419} {\bibfield  {journal} {\bibinfo  {journal} {Phys. Rev. B}\ }\textbf {\bibinfo {volume} {47}},\ \bibinfo {pages} {16419--16426} (\bibinfo {year} {1993})}\BibitemShut {NoStop}%
\bibitem [{\citenamefont {Cheng}\ and\ \citenamefont {Seiberg}(2026)}]{ChengSeiberg2026Peeling}%
  \BibitemOpen
  \bibfield  {author} {\bibinfo {author} {\bibfnamefont {Meng}\ \bibnamefont {Cheng}}\ and\ \bibinfo {author} {\bibfnamefont {Nathan}\ \bibnamefont {Seiberg}},\ }\bibfield  {title} {\enquote {\bibinfo {title} {Peeling off abelian anyons and proliferation transitions},}\ }\href {https://arxiv.org/abs/2610.04704} {\bibfield  {journal} {\bibinfo  {journal} {arXiv e-prints}\ ,\ \bibinfo {eid} {arXiv:2610.04704}} (\bibinfo {year} {2026})},\ \Eprint {http://arxiv.org/abs/2610.04704} {arXiv:2610.04704 [cond-mat.str-el]} \BibitemShut {NoStop}%
\bibitem [{\citenamefont {Gao}\ \emph {et~al.}()\citenamefont {Gao}, \citenamefont {Wang}, \citenamefont {Yang}, \citenamefont {Han},\ and\ \citenamefont {Vishwanath}}]{HanTOSF}%
  \BibitemOpen
  \bibfield  {author} {\bibinfo {author} {\bibfnamefont {Zhi-Qiang}\ \bibnamefont {Gao}}, \bibinfo {author} {\bibfnamefont {Yan-Qi}\ \bibnamefont {Wang}}, \bibinfo {author} {\bibfnamefont {Hui}\ \bibnamefont {Yang}}, \bibinfo {author} {\bibfnamefont {Zhaoyu}\ \bibnamefont {Han}}, \ and\ \bibinfo {author} {\bibfnamefont {Ashvin}\ \bibnamefont {Vishwanath}},\ }\href@noop {} {\enquote {\bibinfo {title} {{SC$^\sharp$: superconductivity intertwined with topological order}},}\ }\bibinfo {note} {To appear}\BibitemShut {NoStop}%
\end{thebibliography}%

\end{document}